\pdfoutput=1
\documentclass[ALICE,manyauthors]{cernphprep}
\newif\ifcernpp
\cernpptrue
\usepackage{hyperref}
\usepackage{lineno}
\usepackage{epstopdf}
\usepackage{graphicx}
\usepackage[loose]{units}
\usepackage{upgreek}
\usepackage{color}
\usepackage[usenames,dvipsnames]{xcolor}
\usepackage{amsmath,amssymb,amsbsy} 
\ifcernpp
\usepackage{cite}
\usepackage[comma,square,numbers,sort&compress]{natbib}
\else
\fi

\begin{document}%

\newcommand{\PbPb}{\textnormal{Pb--Pb}}
\newcommand{\AuAu}{\textnormal{Au--Au}}
\newcommand{\pp}{\ensuremath{\mbox{p}\mbox{p}}}
\newcommand{\pt}{\ensuremath{p_\mathrm{T}}}
\newcommand{\pT}{\pt}
\newcommand{\ptch}{\ensuremath{p_\mathrm{T,\,ch\;jet}}}
\newcommand{\deltaptch}{\ensuremath{\delta p_\mathrm{T,\,ch}}}
\newcommand {\snnbf}{\ensuremath{\mathbf{\sqrt{s_{_{\mathrm{NN}}}} }}}
\newcommand {\snn}{\ensuremath{\sqrt{s_{_{\mathrm{NN}}}} }}

\newcommand{\dirg}       {\ensuremath{{\rm \gamma, dir}}}
\newcommand{\decg}       {\ensuremath{{\rm \gamma, dec}}}
\newcommand{\incg}       {\ensuremath{{\rm \gamma, inc}}}

\newcommand{\Rg}         {\ensuremath{R_{\rm \gamma}}}
\newcommand{\vdirg}      {\ensuremath{v_{2}^{\dirg}}}
\newcommand{\vdecg}      {\ensuremath{v_{2}^{\decg}}}
\newcommand{\vincg}      {\ensuremath{v_{2}^{\incg}}}
\newcommand{\vbckg}      {\ensuremath{v_{2}^{\rm \gamma, bck}}}

\ifcernpp
\begin{titlepage}
\PHyear{2018}
\PHnumber{117}      
\PHdate{9 May}  
\else
\journal{Physics Letters B}
\fi
%
%
\title{Direct photon elliptic flow in Pb--Pb collisions at $\snn=2.76$ TeV}
\ifcernpp
\ShortTitle{Direct photon elliptic flow in Pb--Pb}   
%
\Collaboration{ALICE Collaboration\thanks{See Appendix~\ref{app:collab} for the list of collaboration members}}
\ShortAuthor{ALICE Collaboration} 
\else
\author{ALICE Collaboration}
\fi
\begin{abstract}
The elliptic flow of inclusive and direct photons was measured at mid-rapidity
in two centrality classes 0--20\% and 20--40\% in Pb--Pb collisions at $\snn=2.76$ TeV by ALICE.
Photons were detected with the highly segmented electromagnetic calorimeter PHOS and via conversions in the detector material with the $e^+e^-$ pairs reconstructed in the central tracking system. 
The results of the two methods were combined and the direct-photon elliptic flow was extracted
in the transverse momentum range $0.9 < \pT < \unit[6.2]{GeV}/c$.
A comparison to RHIC data shows a similar magnitude of the measured direct-photon elliptic flow. 
Hydrodynamic and transport model calculations are systematically lower than the data, but are found to be compatible.
\end{abstract}

\ifcernpp
\end{titlepage}
\setcounter{page}{2}
\else
\maketitle
\fi

\section{Introduction}

The theory of the strong interaction, Quantum ChromoDynamics~(QCD), predicts a transition from ordinary hadronic matter to a new state where quarks and gluons are no longer confined to hadrons \cite{Borsanyi:2013lat,Bazavov:2011nk}. 
Lattice calculations predict a chiral and deconfinement crossover transitions over the temperature range 145--163 MeV \cite{Borsanyi:2013lat,Bazavov:2011nk}, which is accessible in collisions of ultrarelativistic heavy ions.
The creation and study of the properties of this hot strongly interacting matter -- Quark-Gluon Plasma (QGP) -- are the main objectives of the ALICE experiment.

The hot strongly interacting matter, created in nucleus--nucleus collisions, expands, cools and finally transforms to ordinary hadronic matter. 
To experimentally study the quark matter properties, several observables were proposed. 
Here, we concentrate on studying the development of {\it collective flow} using {\it direct photons}. 
Direct photons are the photons not originating from hadronic decays but produced in electromagnetic interactions.
Unlike hadrons, direct photons are produced at all stages of the collision. Incoming nuclei passing through each other produce direct photons in scatterings of their partonic constituents. In addition, (thermal) photons are emitted in the deconfined
quark-gluon plasma and hadronic matter, characterized by the thermal
distributions of partons and hadrons, respectively.
Since the mean free path of a photon in hot matter is much larger than the typical sizes of the created fireball \cite{Thoma:1995}, direct photons escape the collision zone unaffected, delivering direct information on the conditions at the production time and on the development of collective flow.

The observations of a strong azimuthal asymmetry of particle production over a wide rapidity range 
in nucleus-nucleus collisions was one of the key results obtained at RHIC \cite{Arsene:2004fa,Back:2004je,Adcox:2004mh,Adams:2005dq} and LHC \cite{Aamodt:2010pa,Alicev2:2011,Aamodt:2011flow,ATLAS:2011ah,Chatrchyan:2012wg} energies.
It was interpreted as a consequence of collective expansion -- collective flow -- of the matter having an initial spatial asymmetry, which is more prominent in collisions with non-zero impact parameter.
To quantify the collective flow, the azimuthal distributions of final state particles are expanded in the series $1+2 \sum v_{n} \cos[n (\varphi-\Psi_{RP})]$ \cite{Voloshin:1994ph}, depending on the difference between the particle azimuthal angle $\varphi$ and the reaction plane orientation $\Psi_{RP}$, defined by the impact parameter and beam axis.
At mid-rapidity the second harmonic $v_2$ (elliptic flow) reflects the expansion of the almond-like shape of the hot matter created by the mutual penetration of the colliding nuclei. 
Higher harmonics $v_3$, $v_4$, etc. are sensitive to fluctuations of the initial shape of the created hot matter and are typically much smaller than $v_2$, except for central collisions, where $v_2$ decreases due to a more symmetric geometry. 
Collective flow is sensitive to the equation of state of hot matter and the amount of shear viscosity. The initial spatial asymmetry of the expanding fireball diminishes with time, for any equation of state. For strongly interacting matter this asymmetry translates into an azimuthal anisotropy in momentum space, while for free streaming weakly interacting matter there is no final particle azimuthal anisotropy.


Hadrons provide the possibility to test with high precision the flow pattern of the latest stage of the collision, when the hot matter decouples into final particles. 
Complementary to them, direct photons provide the possibility to investigate the development of flow during the evolution of hot matter. 
%
%
First calculations predicted that the photon emission rate from the hot quark-gluon or hadron matter increases with temperature as
 $\propto T^2\,\exp(-E_\gamma/T)$ \cite{Kapusta:1991qp}, where $E_\gamma$ is the photon energy and $T$ is the temperature of the matter. Then, the low  transverse momentum ($\pT \lesssim 4$ GeV$/c$) part is controlled by the cooler latest stage, and the high $\pT$ part ($\pT \gtrsim 5$ GeV$/c$) of the spectrum by the hot initial state.
However, detailed calculations which include the full hydrodynamic evolution (see e.g. \cite{Shen:2013vja}) show that 
contributions of all stages are comparable for all $\pT$ regions as a higher temperature of the initial stage is compensated by a larger space-time volume and stronger radial flow of the later stages.
Since the observed direct-photon flow is the convolution of all stages of the collision, including the contribution from the initial stage  when the flow pattern has not yet developed, 
the calculations predict much smaller azimuthal anisotropy for thermal photons than for hadrons \cite{Chatterjee:2013naa, Shen:2013cca}.


~\\
The first measurement of a direct-photon spectrum in relativistic nucleus-nucleus collisions was presented by the WA98 collaboration \cite{WA98:2000}, and later also by the PHENIX Collaboration \cite{PHENIX:2005a,PHENIX:2010a,PHENIX:2012a,PHENIX:2015a}, and by the ALICE Collaboration \cite{Adam:2015lda}. The first measurement of elliptic flow of direct photons in Au--Au collisions at $\snn=200$ GeV was performed by the PHENIX Collaboration \cite{Adare:2011zr}. 
Surprisingly, it was found to be close to the flow of hadrons \cite{Adare:2011tg}. 
Recent PHENIX results, presenting more precise measurements of elliptic and triangular flow extended to lower $\pT$ \cite{Adare:2015lcd}, confirmed this early result. 
The discrepancy between experimental results and theory predictions triggered a set of theoretical studies, which can be split into two classes. The main idea in the first class of models \cite{Chatterjee:2005de,Chatterjee:2008tp,vanHees:2011vb,Linnyk:2013wma,Gale:2014dfa,Muller:2013ila,vanHees:2014ida,Monnai:2014kqa,Dion:2011pp,Liu:2012ax,Vujanovic:2014xva,McLerran:2014hza,McLerran:2015mda,Gelis:2004ep,Hidaka:2015ima,Linnyk:2015rco,Vovchenko:2016ijt,Koide:2016kpe} is to increase the emission of direct photons from the later stages of the collision and/or suppress emission of the initial stage. 
In the second class of models \cite{Turbide:2005bz,Tuchin:2012mf,Basar:2012bp}, a new azimuthally asymmetric source of direct photons is considered like jet-matter interactions or synchrotron radiation in the field of colliding nuclei. These theoretical efforts considerably reduce the discrepancy, but consistent reproduction of both the direct-photon spectra and flow is still missing. The measurement of direct-photon flow at higher collision energy is important as an independent confirmation of the results at lower energy, and could also allow to disentangle between different contributions.

In this paper, we present the first measurement of the direct-photon flow in Pb--Pb collisions at the LHC 
and compare our findings to RHIC results and to predictions of 
hydrodynamic as well as transport models.

\section{Detector setup}

The direct photon flow is based on the measurement of the elliptic flow of inclusive photons and the estimation of the contribution of decay photons using the available hadron flow results.
Photons are reconstructed via two independent methods: the Photon Conversion Method (PCM) and with the electromagnetic calorimeter PHOS.

In the conversion method, the electron and positron tracks from photon conversions are measured with the Inner Tracking System (ITS) and/or the Time Projection Chamber (TPC). The ITS \cite{Aamodt:2008zz} consists of two layers of Silicon Pixel Detectors (SPD) positioned at radial distances of \unit[3.9]{cm} and \unit[7.6]{cm}, two layers of Silicon Drift Detectors (SDD) at \unit[15.0]{cm} and \unit[23.9]{cm}, and two layers of Silicon Strip Detectors (SSD) at \unit[38.0]{cm} and \unit[43.0]{cm}. 
The two innermost layers cover a pseudorapidity range of $|\eta|<2$ and $|\eta|<1.4$, respectively.
The TPC \cite{Alme:2010ke} is a large (85~m$^3$) cylindrical drift detector filled with a Ne-CO$_2$-N$_2$ (85.7/9.5/4.8\%) gas mixture. It covers the pseudorapidity range $|\eta|<0.9$ over the full azimuthal angle with a maximum track length of 159 reconstructed space points. With the solenoidal magnetic field of $B=\unit[0.5]{T}$, electron and positron tracks can be reconstructed down to $\pT \approx \unit[50]{MeV}/c$. The TPC provides particle identification via the measurement of the specific energy loss (d$E$/d$x$) with a resolution of 5.2\% in pp collisions and 6.5\% in central Pb-Pb collisions \cite{Abelev:2014ffa}. The ITS and the TPC were aligned with respect to each other to the level of less than $\unit[100]{\mathit{\mu} m}$ using cosmic-ray and \pp\ collision data \cite{Aamodt:2010aa}. Particle identification is also provided by the Time-of-Flight (TOF) detector \cite{Akindinov:2009zze} located at a radial distance of $370 < r < \unit[399]{cm}$. This detector consists of Multigap Resistive Plate 
Chambers (MRPC) and provides timing information with an intrinsic resolution of \unit[50]{ps}.

PHOS \cite{Dellacasa:1999kd} is an electromagnetic calorimeter which consists of three modules installed at a radial distance of \unit[4.6]{m} from the interaction point. It subtends $260^\circ<\varphi<320^\circ$ in azimuth and $|\eta|<0.12$ in pseudorapidity. Each module consists of 3584 detector cells arranged in a matrix of $64\times 56$ lead tungstate crystals each of size \unit[$2.2\times 2.2 \times 18$]{cm$^3$}. The signal from each cell is measured by an avalanche photodiode (APD) associated with a low-noise charge-sensitive preamplifier. To increase the light yield, reduce electronic noise, and improve energy resolution, the APDs and preamplifiers are cooled to a temperature of $-25~^\circ$C. The resulting energy resolution is $\sigma_{E}/E=(1.8\%/E) \oplus (3.3\%/\sqrt{E})\oplus 1.1$\%, where $E$ is in units of GeV. The energy deposition in each PHOS cell is calibrated in \pp\ collisions by aligning the $\pi^0$ peak position in the two-photon invariant mass distribution.

For the minimum bias trigger in the \PbPb\ run and event plane orientation calculation, two scintillator array detectors (V0--A and V0--C) \cite{Cortese:2004aa} are used, which subtend $2.8 < \eta < 5.1$ and $-3.7 < \eta < -1.7$, respectively. 
Each of the V0~arrays consists of 32 channels and is segmented
in four rings in the radial direction, and each ring is divided into eight sectors in the azimuthal direction. The sum of the signal amplitudes of the V0--A and V0--C detectors serves as a measure of centrality in the \PbPb\ collisions.

\section{Data analysis}

This analysis is based on data recorded by the ALICE experiment in the first LHC heavy-ion run in the fall of 2010. The detector readout was triggered by the minimum bias interaction trigger based on signals from the V0--A, V0--C, and SPD detectors. The efficiency for triggering on a \PbPb\ hadronic interaction ranged between 98.4\% and 99.7\%. The events are divided into the central and semi-central centrality classes 0--20\% and 20--40\%, respectively, according to the V0--A and V0--C summed amplitudes \cite{Abelev:2013qoq}. 
To ensure a uniform track acceptance in pseudorapidity $\eta$, only events with a primary vertex within $\pm \unit[10]{cm}$ from the nominal interaction point along the beam line ($z$-direction) are used. After offline event selection, $13.6 \times 10^6$ events are available for the PCM analysis and $18.8 \times 10^6$ events for the PHOS analysis. 

The direct-photon elliptic flow is extracted on a statistical basis by subtracting the elliptic flow of photons from hadron decays from the inclusive photon elliptic flow. We assume that in each bin of the photon transverse momentum the measured inclusive photon flow can be decomposed as
\begin{equation}
  \vincg = \frac{N_{\dirg}}{N_{\incg}} \vdirg + \frac{N_{\decg}}{N_{\incg}}\vdecg,
\end{equation}
where $N_{\incg}=N_{\dirg}+N_{\decg}$ is the inclusive photon yield which can be decomposed into the contributions of direct ($N_{\dirg}$) and decay ($N_{\decg}$) photons. The $\vincg$, $\vdirg$ and $\vdecg$ are the corresponding photon flows. It is convenient to express direct-photon flow in terms of the ratio $\Rg=N_{\incg}/N_{\decg}$, the inclusive photon flow $\vincg$, and the decay photon flow $\vdecg$:
\begin{equation}
  \vdirg=\frac{\vincg \Rg - \vdecg }{\Rg-1}.
  \label{vdir}
\end{equation}
The ratio $\Rg$ was measured in the same dataset in \cite{Adam:2015lda}, whereas $\vdecg$ is calculated with a simulation of photons from decays which is also known as cocktail simulation.
The PCM and PHOS measurements of inclusive photon flow are performed independently. They are then combined and used with the combined ratio $\Rg$ as well as the calculated decay photon flow. 

The photon elliptic flow $v_2$ is calculated with the Scalar Product (SP) method, which is a two-particle correlation method \cite{Adler:2002pu}, 
using a pseudorapidity gap of $|\Delta\eta| > 0.9$ between the photon and the reference flow particles. The applied gap reduces correlations not related to the event plane $\Psi_n$, such as the ones due to resonance decays and jets, known as non--flow effects.
The SP method uses the $Q$-vector, computed from a set of reference
flow particles (RFP) defined as:
\begin{equation}
  \vec{Q}_n = \sum_{i \in \mathrm{RFP}} w_{i} e^{in\varphi_{i}},
\label{eq:qvector}
\end{equation}
where $\varphi_i$ is the azimuthal angle of the $i$-th RFP, $n$ is the order of the harmonic and $w_i$ is a weight applied for every RFP. The RFPs are taken from the V0--A and V0--C detectors.
Since these detectors do not provide tracking information,
we sum over the V0--A/V0--C cells, 
while the 
amplitudes of the signal from each cell, which are proportional to the number of particles that cause a hit, are used as a weight $w_i$. 
The non-uniformity of the detector azimuthal efficiency is taken into account 
by applying the inverse of the event-averaged signal as a weight for each of the V0~segments, 
together with a recentering procedure~\cite{Abelev:2014ffa,Abbas:2013taa}. More specifically, the elliptic flow $v_2$ is calculated using the unit flow vector 
$\vec{\mathrm{u}}_2 =  e^{i2\varphi}$ built from reconstructed photons 

\begin{equation}
 v_2 = 
\sqrt{
  \frac{
  \big\langle \big\langle \vec{\mathrm{u}}_2 \cdot \frac{\vec{Q}_2^\mathrm{A*}}{M_\mathrm{A}}\big\rangle \big\rangle
  \big\langle \big\langle \vec{\mathrm{u}}_2 \cdot \frac{\vec{Q}_2^{\mathrm{C*}}}{M_{\mathrm{C}}}\big\rangle \big\rangle
  }{
  \big\langle \frac{\vec{Q}_2^\mathrm{A}}{M_\mathrm{A}} \cdot \frac{\vec{Q}_2^\mathrm{C*}}{M_\mathrm{C}} \big\rangle
  }
},
\label{eq:sp}
\end{equation}
where the two pairs of brackets in the numerator indicate an average over all photons and over all events;
$M_{\mathrm{A}}$ and $M_{\mathrm{C}}$ are the estimates of multiplicity from the V0--A and V0--C detectors, respectively; 
and $\vec{Q}_2^\mathrm{A*}$, $\vec{Q}_2^\mathrm{C*}$ are the complex conjugates of the flow vector calculated in sub-event A and C, 
respectively.

In the PCM analysis, photons converting into e$^+$e$^-$ pairs are reconstructed with an algorithm which searches for displaced vertices with two oppositely charged daughter tracks. Only good quality TPC tracks with a transverse momentum above \unit[50]{MeV}/$c$ and a pseudorapidity  of $\vert \eta \vert < 0.9$ are considered.
The vertex finding algorithm uses the Kalman filter technique for the decay/conversion point and four-momentum determination of the neutral parent particle ($V^0$) \cite{Abelev:2012cn}. Further selection is performed on the level of the reconstructed $V^0$. Only $V^0$s with a conversion points at radii between $5<R<\unit[180]{cm}$ are accepted such that the $\pi^0$ and $\eta$-meson Dalitz decays are rejected and to ensure a good coverage by the tracking detectors of the conversion daughters. To identify an e$^+$e$^-$ pair, the specific energy loss~(d$E$/d$x$) in the TPC \cite{Abelev:2014ffa} of both daughters is used. The transverse momentum component $q_T$ 
of the electron momentum, $p_e$, with respect to the $V^0$ momentum-vector is restricted to be $q_T < 0.05 \sqrt{1- \left( \alpha / 0.95\right)^{2}}$~GeV/$c$, where $\alpha$ is the energy asymmetry of the conversion daughters. Random associations of electrons and positrons are further reduced by selecting $V^0$s with $\mathrm{cos}(\theta)>0.85$, where $\theta$ is the pointing angle, which is the angle between the momentum-vector of the $e^+e^-$ pair and the vector that connects the primary vertex and the conversion point. Based on the invariant mass of the $e^{+} e^{-}$ pair and the pointing angle of the $V^0$ to the primary vertex, the vertex finder calculates a $\chi^2$ value which reflects the level of consistency with the hypothesis that the $V^0$ comes from a photon originating from the primary vertex. A selection based on this $\chi^2$ value is used to further reduce contamination in the photon sample.
The main sources of background that remain after these selection criteria are V0s reconstructed from $\pi^{\pm} \mathrm{e}^{\mp}$, $\pi^{\pm} \pi^{\mp}$, $\pi^{\pm} K^{\mp}$ and e$^{\pm} K^{\mp}$ pairs, which is important to take into account as shown in \cite{Bock:2016ex}. The elliptic flow of this background is subtracted using a side-band method approach. In this method, the d$E$/d$x$ information of both conversion daughters is combined into a 1-dimensional quantity. The signal is a peaked distribution and the side-bands are dominated by background sources. The $v_2$ of the side-bands is measured and subtracted from the main signal region using the purity of the photon sample, which is obtained by fitting Monte Carlo templates to the data. The correction to the measured inclusive photon flow is of the order of 5\% for central and 2.5\% for semi-central collisions, respectively.



The systematic uncertainties of the inclusive photon flow measured with PCM are summarized in Tab.~\ref{tab:sys_inc}. 
The uncertainties related to the photon selection ($|\eta|$, $R$, min $\pT$, $q_{\mathrm{T}}$, $\chi^{2}/ \mathrm{ndf}$ and  
$\cos(\theta)$) are obtained by varying the selection criteria, and the systematic uncertainties related to the contamination of the photon sample are quantified by the uncertainty on the background flow subtraction. The energy resolution uncertainties, which are due to detector resolution effects and bremsstrahlung of electrons, are estimated by comparing $\vincg$ distributions as a function of the reconstructed and true $\pT$ using MC simulations.
The uncertainties related to the variation of reconstruction efficiency in- and out-of-plane are calculated from studying the photon reconstruction efficiency as a function of the track multiplicity. For most of these sources only a small dependence on $\pT$ and collision centrality is observed.

\begin{table}[t]
\centering
\begin{tabular}{lcccc}
     \hline
Centrality     & \multicolumn{2}{c}{0--20\%} & \multicolumn{2}{c}{20--40\%}  \\
\hline
$\pT$ (GeV/$c$)               												& 2.0 & 5.0 & 2.0 & 5.0 \\
\hline
\hline
\textbf{PCM} & & & & \\
Photon selection													& 2.4 & 4.2 & 2.1 & 4.0 \\
Energy resolution        											& 1.0 & 1.0 & 1.0 & 1.0 \\
Efficiency										& 3 & 3 & 1.9 & 1.9 \\
\hline
Total     		& 4.0 & 5.3 & 3.0 & 4.5\\
\hline
\hline
\textbf{PHOS} & & & & \\
Efficiency \& contamination    &  3.0  & 3.0 & 0.7 & 0.7 \\
Event plane flatness &  2.0  & 2.0 & 1.4 & 1.4 \\
\hline
Total  &3.5 & 3.5  &1.6 & 1.6 \\
\hline
\hline
\textbf{Decay photon calculation} & & & & \\
Parameterization of $v_{2}^\pi$   &  1.3  & 3.6 & 0.8 & 2.2 \\
$\eta/\pi^0$ normalization        &  1.7  & 3.2 & 1.7 & 2.4 \\
\hline
Total 		      &  2.2  & 4.8 & 1.9 & 3.3 \\
\hline
\hline
\end{tabular} 
\caption{\label{tab:sys_inc}
Summary of the relative systematic uncertainties (in \%) of the inclusive photon elliptic flow in the PCM and PHOS analysis, and of the decay photon simulation. 
All contributions are expected to be correlated in $\pT $ with the magnitude of the relative uncertainty varying point-by-point.}
\end{table}

In the PHOS analysis, the same photon selection criteria are applied as in the direct-photon spectra analysis \cite{Adam:2015lda}. Cells with a common edge with another cell that are both above the energy threshold of 25 MeV are combined into clusters which are used as photon candidates. 
To estimate the photon energy, the energies of all cells or only those with centers within a radius $R_{\rm core} = \unit[3.5]{cm}$ from the cluster center of gravity are summed.  Compared to the full cluster energy, the core energy is less sensitive to overlaps with low-energy clusters in a high multiplicity environment, and is well reproduced by GEANT3 Monte Carlo simulations \cite{Adam:2015lda}.The full energy is used for the systematic uncertainty estimate.
The contribution of hadronic clusters is reduced by requiring $E_{\rm cluster}>\unit[0.3]{GeV}$, $N_{\mathrm{cells}}>2$ and by accepting only clusters above a minimum lateral cluster dispersion \cite{Abelev:2014ypa}. The latter selection rejects rare events when hadrons punch through the crystal and 
hadronically interact with APD, producing a large signal in one cell of a cluster, not proportional to the energy deposition.
In addition to these cuts, we also apply 
a $\pT$-dependent dispersion cut and perform a charged particle veto (CPV). The CPV removes clusters based on the minimal distance between the PHOS cluster position and the position of extrapolated charged tracks on the PHOS surface, and is used to suppress hadron contribution \cite{Abelev:2014ypa}. Both dispersion and CPV cuts are tuned using \pp\ collision data to provide the photon reconstruction and identification efficiency at the level of 96--99\%. Measurements with different combinations of dispersion and CPV cuts are used for the
estimate of systematic uncertainties.
Possible pileup contribution from other bunch crossings is removed by a loose cut on the cluster arrival time $|t|<\unit[150]{ns}$, which is small compared to a minimum time between bunch crossings of \unit[525]{ns}. 

To estimate the reconstruction and identification efficiencies and correction for energy smearing with their possible dependence on the angle with respect to the event plane, 
we embedded simulated photon clusters into real data events and applied the standard reconstruction procedure. PHOS properties (energy and position resolutions, residual de-calibration, absolute calibration, non-linear energy response) are tuned in the simulation to reproduce the $\pT$-dependence of the $\pi^0$ peak position and width \cite{Abelev:2014ypa}. The correction for the event plane dependence of the reconstruction and identification efficiencies, which comes as additive to the observed photon flow, is less than $10^{-3}$ both in central and mid-central collisions and is comparable to the statistical uncertainties of the embedding procedure. 
The correction due to the energy smearing, is estimated to be 4\% and 1\% for central and semi-central collisions, respectively.   
The contamination of the photon sample measured with PHOS originates mainly from $\pi^{\pm}$ and $\bar{p}$, $\bar{n}$ annihilation, with other contributions being much smaller. The application of the dispersion and CPV cuts reduces the overall contamination at $\pT \approx 1.5 $ GeV/$c$ from about 15\% to 2--3\% and down to 1--2\% at $\pT\sim 3$--$\unit[4]{GeV}/c$. 
To estimate and subtract the hadron contribution, the PHOS response matrices are constructed for $\pi^\pm$, 
K, p and $\bar{p}$ using real data or Monte Carlo simulations and convoluted with the measured spectra, flow and relative yields of hadrons. 

Systematic uncertainties of the inclusive photon flow measured with PHOS are summarized in Tab.\ \ref{tab:sys_inc}. 
They can be split into two groups: contributions related to the contamination and dependence of reconstruction, identification and smearing efficiency on the angle with respect to the event plane, and uncertainties related to the flatness of the event plane calculation, the event plane resolution and the contribution of non-flow effects. Uncertainties of the first group are estimated by comparing the fully corrected photon flow measured with different sets of identification criteria and with full and core energy. Uncertainties of the second group are estimated by comparing 
inclusive photon flows measured separately with the V0--A and V0--C detectors. Note that because of the limited azimuthal acceptance, PHOS is much more sensitive to the non-flatness of the event plane distribution compared to PCM.

\begin{figure}[ht]
\unitlength\textwidth
\centering
\includegraphics[width=0.48\linewidth]{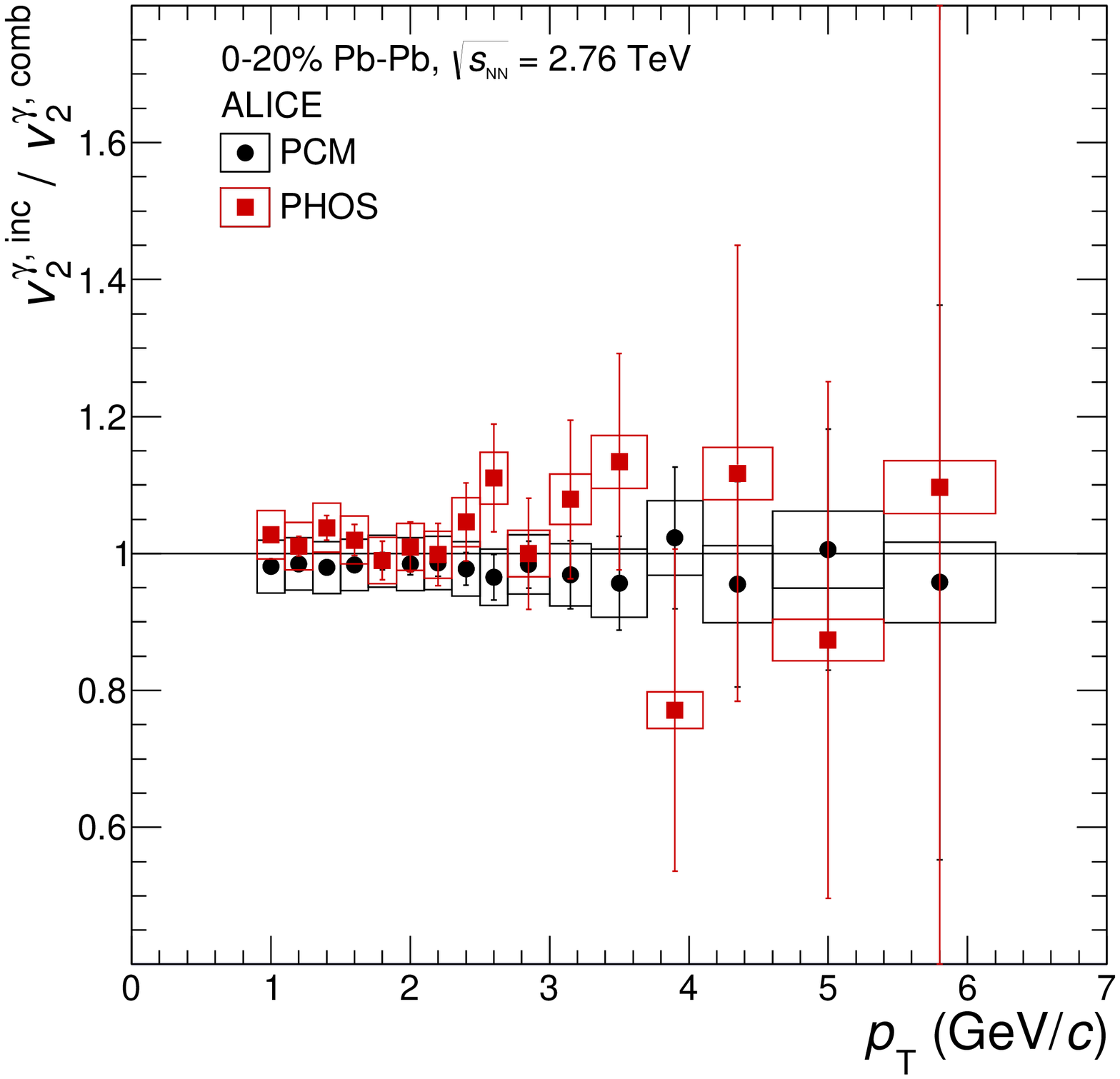}
\hfill
\includegraphics[width=0.48\linewidth]{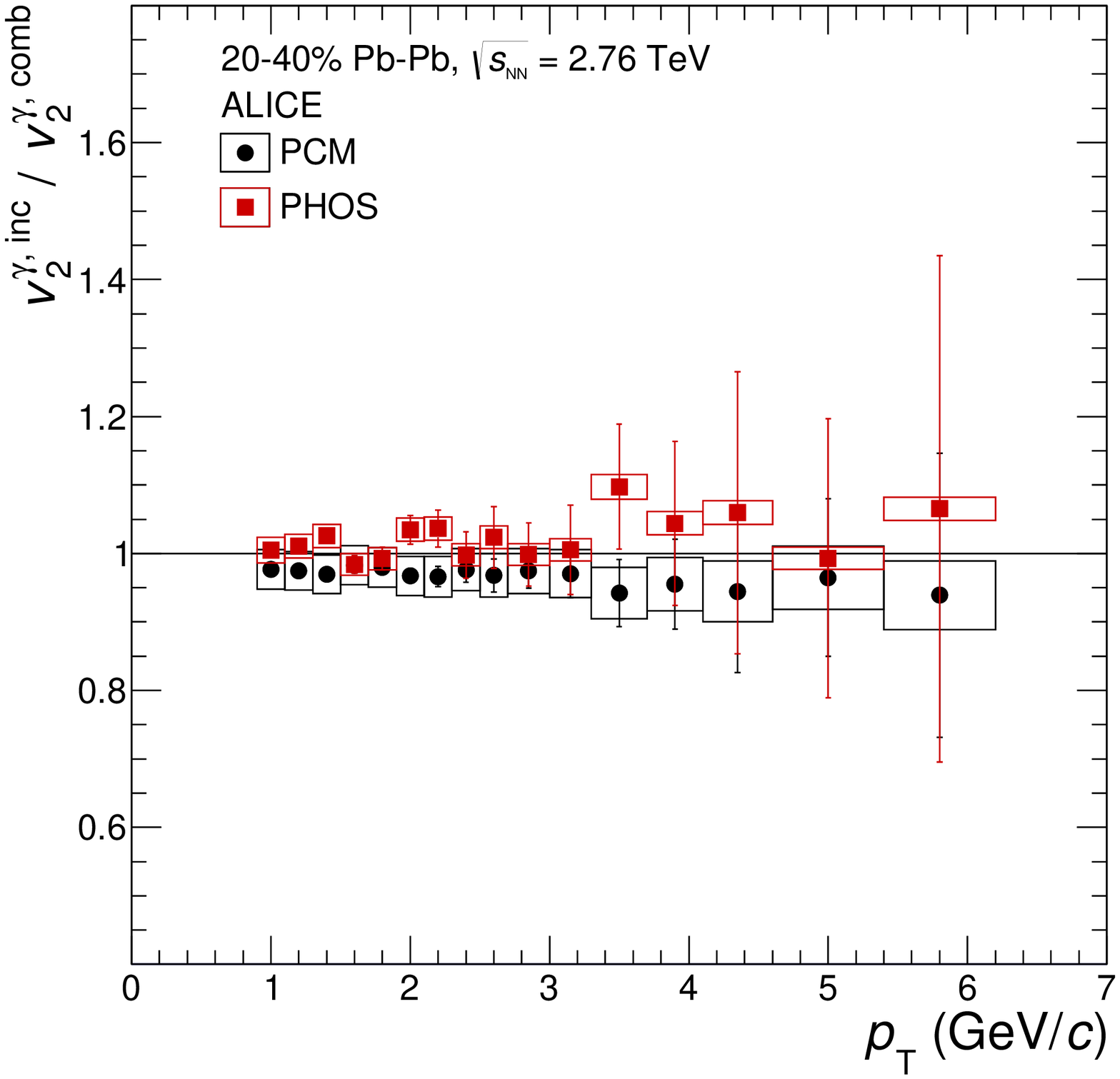}
\caption{
\label{fig:Comp-PHOS-PCM} 
(Color online) Comparison of the measured inclusive photon flow $(\vincg)$ 
to the individual PCM and PHOS measurements $(v_{2}^{\rm \gamma, ind})$ in the 0--20\% (left) and 20--40\% (right) centrality classes. The individual results are divided by the combined $\vincg$. The vertical bars on each data point indicate the statistical uncertainties and the boxes indicate the systematic uncertainties.
}
\end{figure}

In the combination of the inclusive photon $v_2$ results from PCM and PHOS, both measurements are treated as independent. Possible correlations due to the use of the same V0A and V0C event plane vectors are found to be negligible. To take into account correlations of the individual measurements in bins of transverse momentum, we describe the measured inclusive photon flows as vectors $\vec v_2^{\gamma,\mathrm{inc,PCM}}$, $\vec v_2^{\gamma,\mathrm{inc,PHOS}}$, where the vector components correspond to the measured $\pT$ bins, and the correlations of the total uncertainties are described by covariance matrices $V_{v_2,\mathrm{PCM}}$ and $V_{v_2,\mathrm{PHOS}}$, respectively. The elements of the covariance matrix are calculated assuming uncorrelated statistical uncertainties and fully correlated $(\rho = 1)$ systematic uncertainties; $V_{ij} = V_{\mathrm{stat},ij} + V_{\mathrm{syst},ij}$, where $V_{\mathrm{syst},ij} = \rho \sigma_{\mathrm{syst},i} \sigma_{\mathrm{syst},j}$, for $\pT$ bin $i$ and $j$. Then, the combined inclusive photon flow is the vector 
\begin{equation}
\vec v_2^{\gamma,\mathrm{inc}} = (V_{v_2,\mathrm{PCM}}^{-1} + V_{v_2,\mathrm{PHOS}}^{-1})^{-1} (V_{v_2,\mathrm{PCM}}^{-1} \vec v_2^{\gamma,\mathrm{inc,PCM}} + V_{v_2,\mathrm{PHOS}}^{-1} \vec v_2^{\gamma,\mathrm{inc,PHOS}}) \label{v2incl}.
\end{equation}
The inclusive photon $v_2$ measured with PCM and PHOS are compared in 
Fig.\ \ref{fig:Comp-PHOS-PCM}, which shows the ratio of the individual values to the combined flow. The PCM and PHOS measurements are found to be consistent with each other with $p$-values of 0.93 and 0.43 for the centrality classes 0--20\% and 20--40\%, respectively.

\begin{figure}[ht]
\unitlength\textwidth
\centering
\includegraphics[width=0.48\linewidth]{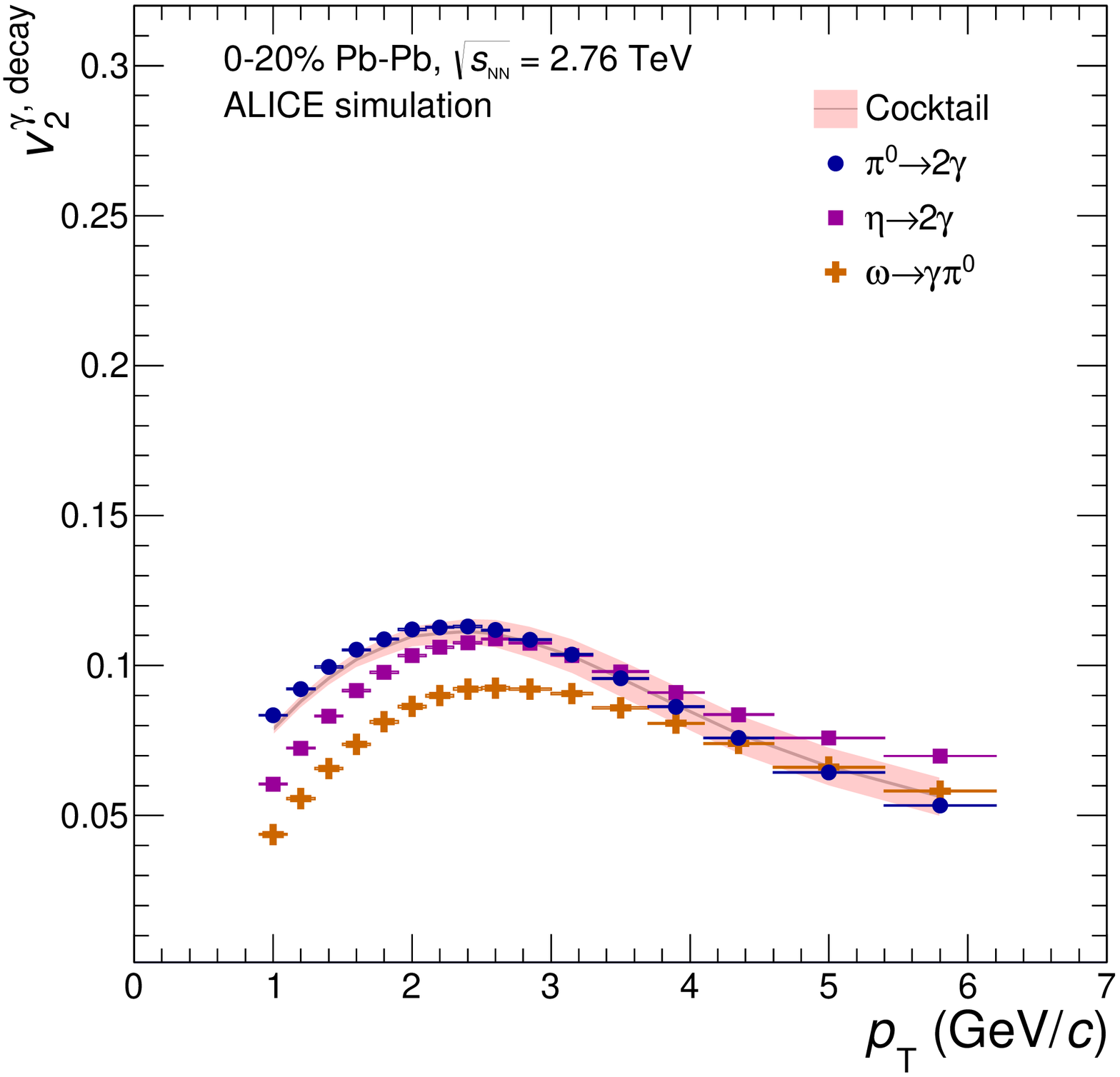}
\hfill
\includegraphics[width=0.48\linewidth]{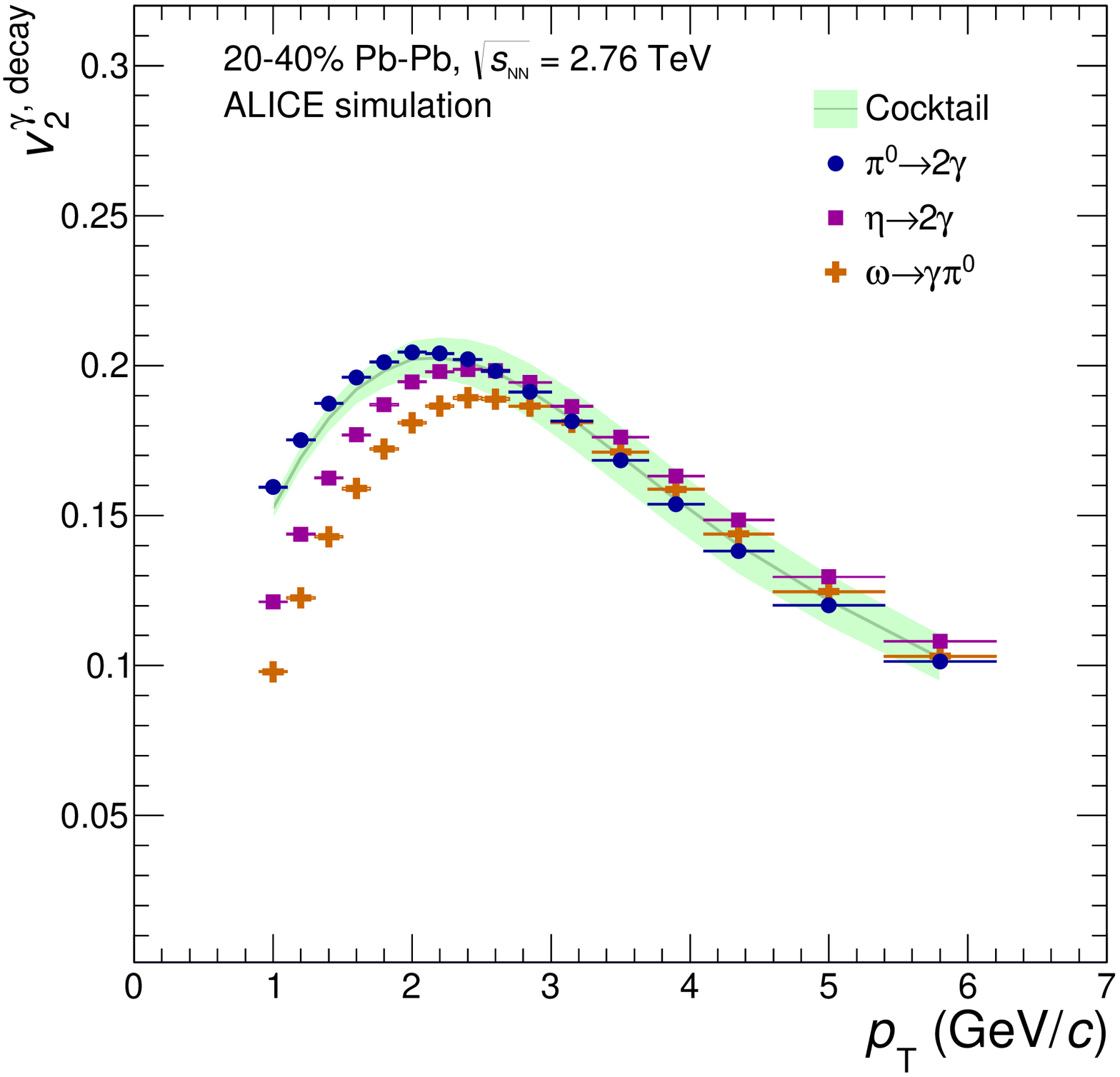}
\caption{
\label{fig:Cocktail} 
(Color online) Elliptic flow of decay photons from $\pi^{0}$, $\eta$, $\omega$, and the total cocktail simulation as a function of transverse momentum in the 0--20\% (left) and 20--40\% (right) centrality classes. The band represents the total uncertainty of the total cocktail simulation.}
\end{figure}

The decay photon flow is estimated using a cocktail simulation.
Decays that contribute more than 1\% of the total decay photon yield are taken into account:
$\pi^0\to2\gamma$, $\eta\to 2\gamma$, $\omega \to \gamma\pi^0$, $K_{s}^{0}\to2\pi^0\to 4\gamma$. Other contributions are negligible compared to the systematic uncertainties of the cocktail. In decays of $\eta$ and $\omega$ mesons only photons produced directly in decays are accounted, 
while those coming from daughter $\pi^0$ decays are already accounted in $\pi^0$ contribution.
The $K_s^0$ decay does not contribute significantly to the photon sample measured with the PCM approach. Therefore, we correct the PHOS measurement for this contribution before combining the PHOS and PCM measurements. 
Here we use the same approach as in the direct-photon spectrum analysis \cite{Adam:2015lda}, but this time 
the simulation of the elliptic flow is added. 
To estimate the elliptic flow of neutral pions, a parametrization has been made of the charged pion flow measured under the same conditions, i.e.,~charged pions measured in the TPC and reference particles in the V0--A and V0--C detectors \cite{Abelev:2014pua,Abelev:2012di} are used. To estimate the contribution of $\eta$ and $\omega$ mesons, the measured elliptic flow of charged and neutral kaons \cite{Abelev:2014pua} is scaled, assuming scaling with the transverse kinetic energy $KE_\mathrm{T} = m_\mathrm{T} - m$.
The comparison of different contributions and overall decay photon flow is shown in Fig.\ \ref{fig:Cocktail}. 
The $v_2$ contributions were added with weights, proportional to the relative decay photon yield of a meson in total decay yield \cite{Adam:2015lda}.
The width of the colored band represents the systematic uncertainties of the decay photon elliptic flow $\vdecg$.
The decay photon flow is mainly determined by the $\pi^0$ flow, while other contributions make relatively small corrections:
the $\eta$ and $\omega$ contributions slightly reduce the decay photon elliptic flow at $\pT<2$ GeV/$c$ and increase it compared to the $\pi^0$ contribution at higher $\pT$.
The systematic uncertainties of the decay photon flow are summarized in Tab.\ \ref{tab:sys_inc}. The largest uncertainties come 
from the parametrization of the charged pion elliptic flow and from the relative yield $\eta/\pi^{0}$.

\section{Results}

\begin{figure}[t]
\unitlength\textwidth
\centering
\includegraphics[width=0.49\linewidth]{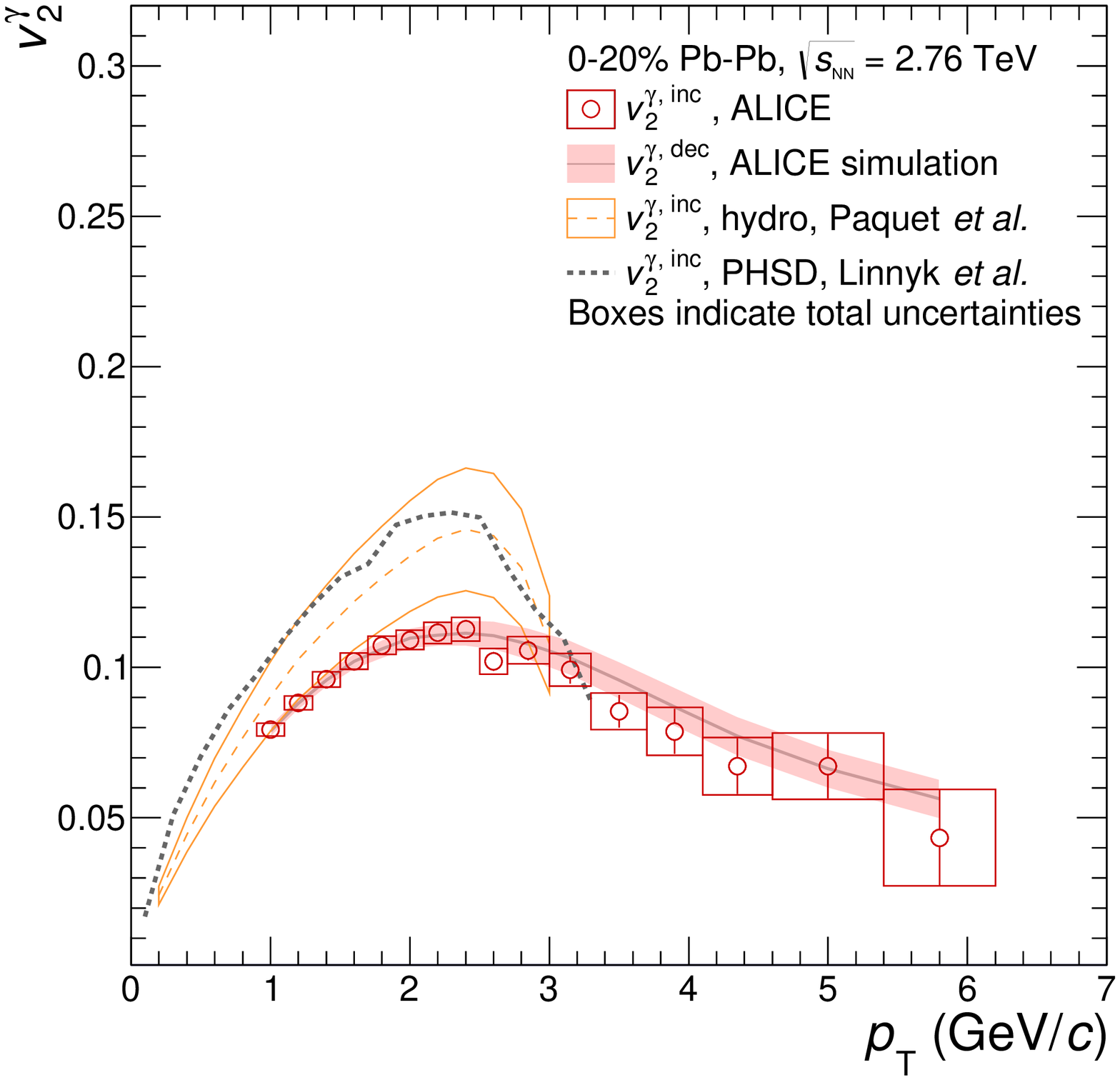}
\hfill
\includegraphics[width=0.49\linewidth]{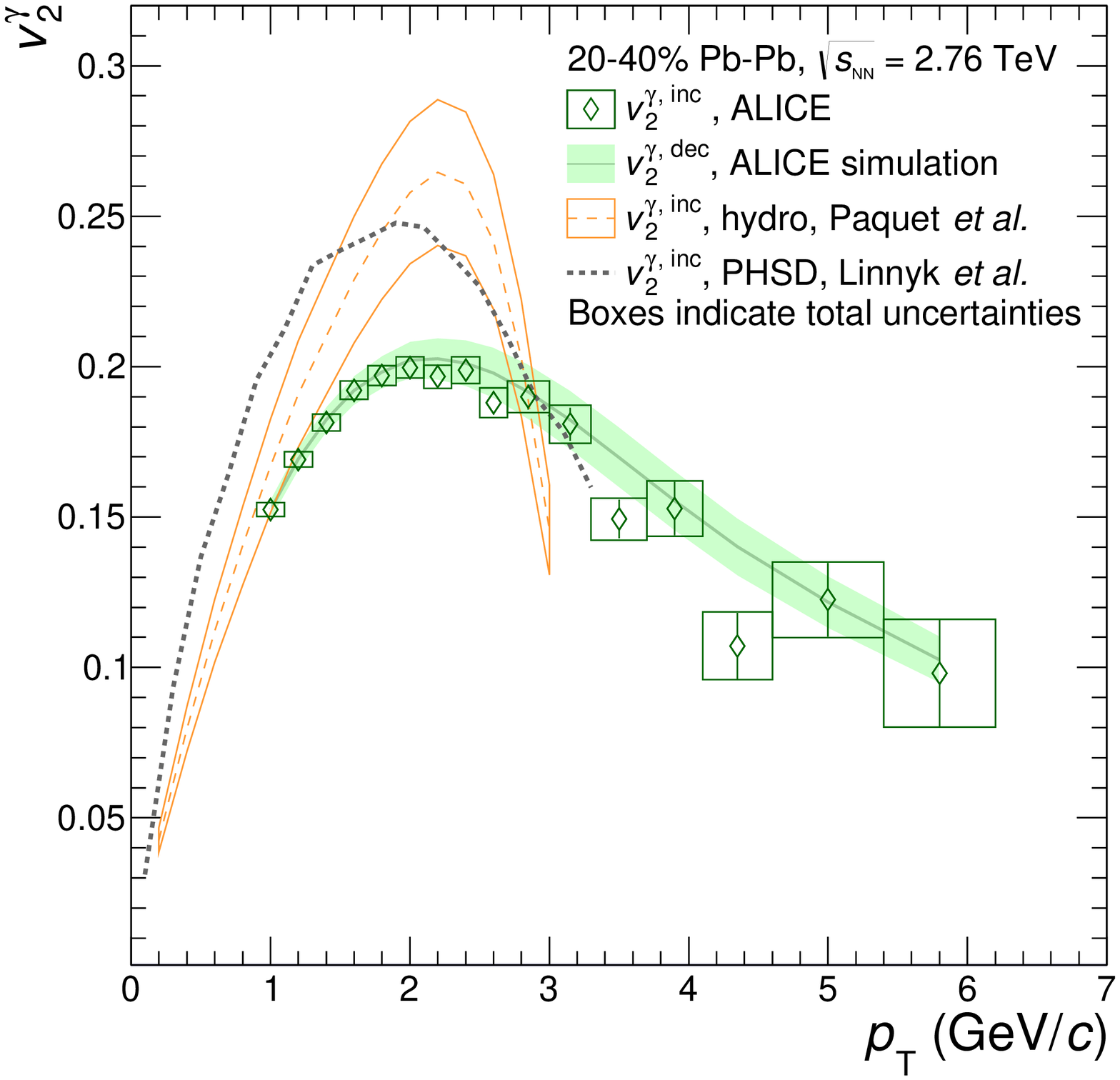}
\caption{(Color online) Elliptic flow of inclusive photons and decay photons, compared to hydrodynamic \cite{Gale:2014dfa} and transport PHSD \cite{Linnyk:2013wma} model predictions in the 0--20\% (left) and 20--40\% (right) centrality classes. The vertical bars on each data point indicate the statistical uncertainties and the boxes indicate the sizes of the total uncertainties.
\label{fig:Inclv2Theory}}
\end{figure}

The $\vincg$ measured in two centrality classes are shown in Fig.~\ref{fig:Inclv2Theory}. 
The elliptic flow coefficients of inclusive photons and decay photons are very similar over the full range $0.9<\pT<6.2$~GeV/$c$. As the fraction of direct photon over the inclusive photon yield is relatively small, $\sim 10$\% in our $\pT$ range \cite{Adam:2015lda}, the collective flow of inclusive photons is dominated by the decay photon flow. In models based on relativistic hydrodynamics the medium is assumed to be in or close to local thermal equilibrium. An equation of state is used to relate thermodynamic quantities like temperature, energy density, and pressure. Photon production is modeled by folding the space-time evolution of a collision with temperature-dependent photon production rates in the QGP and the hadron gas. Another approach is taken, e.g., in the PHSD transport model in which the QGP degrees of freedom are modeled as massive strongly-interacting quasi-particles \cite{Linnyk:2015tha}. For both classes of models the development of a strong early elliptic flow, necessary to reproduce the observed direct-photon flow, gives rise to a large pion elliptic flow at freeze-out and therefore to a large inclusive photon elliptic flow. It is therefore an important test to check whether a model can describe both the inclusive and the direct-photon elliptic flow. The prediction of the hydrodynamic model described in \cite{Paquet:2016} for the inclusive photon $v_2$ in the range $1<\pT<3$~GeV/$c$ is about 40\% above the data, though the magnitude of the elliptic flow of unidentified hadrons is reproduced within 10--20\% accuracy in this $\pT$ range \cite{Ryu:2017qzn}. The PHSD model \cite{Linnyk:2015tha} also predicts an $\sim$ 40\% higher inclusive photon flow, even though it reproduces the unidentified hadron flow well.

\begin{figure}[ht]
\unitlength\textwidth
\centering
\includegraphics[width=0.50\linewidth]{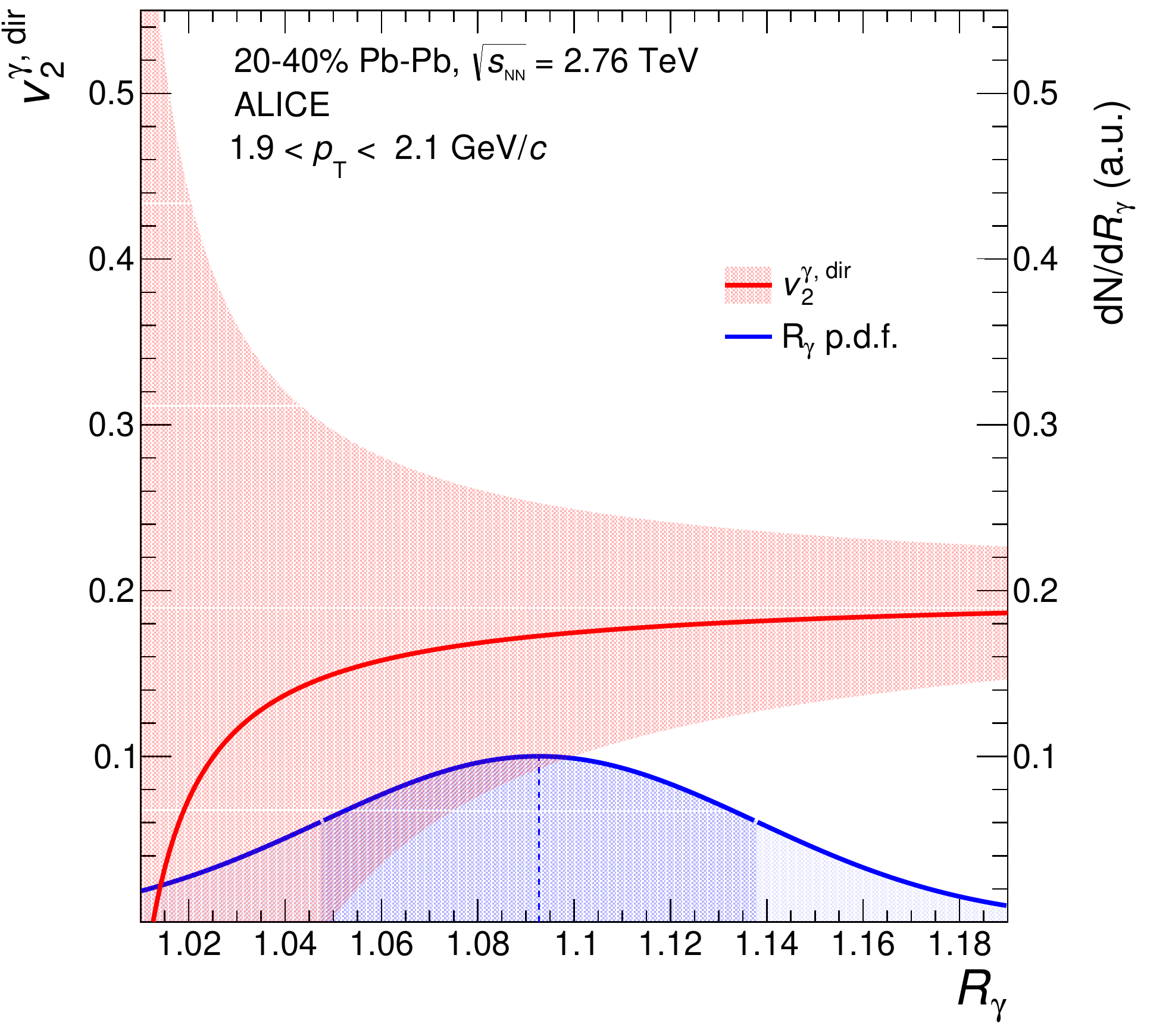}
\hfill
\includegraphics[width=0.455\linewidth]{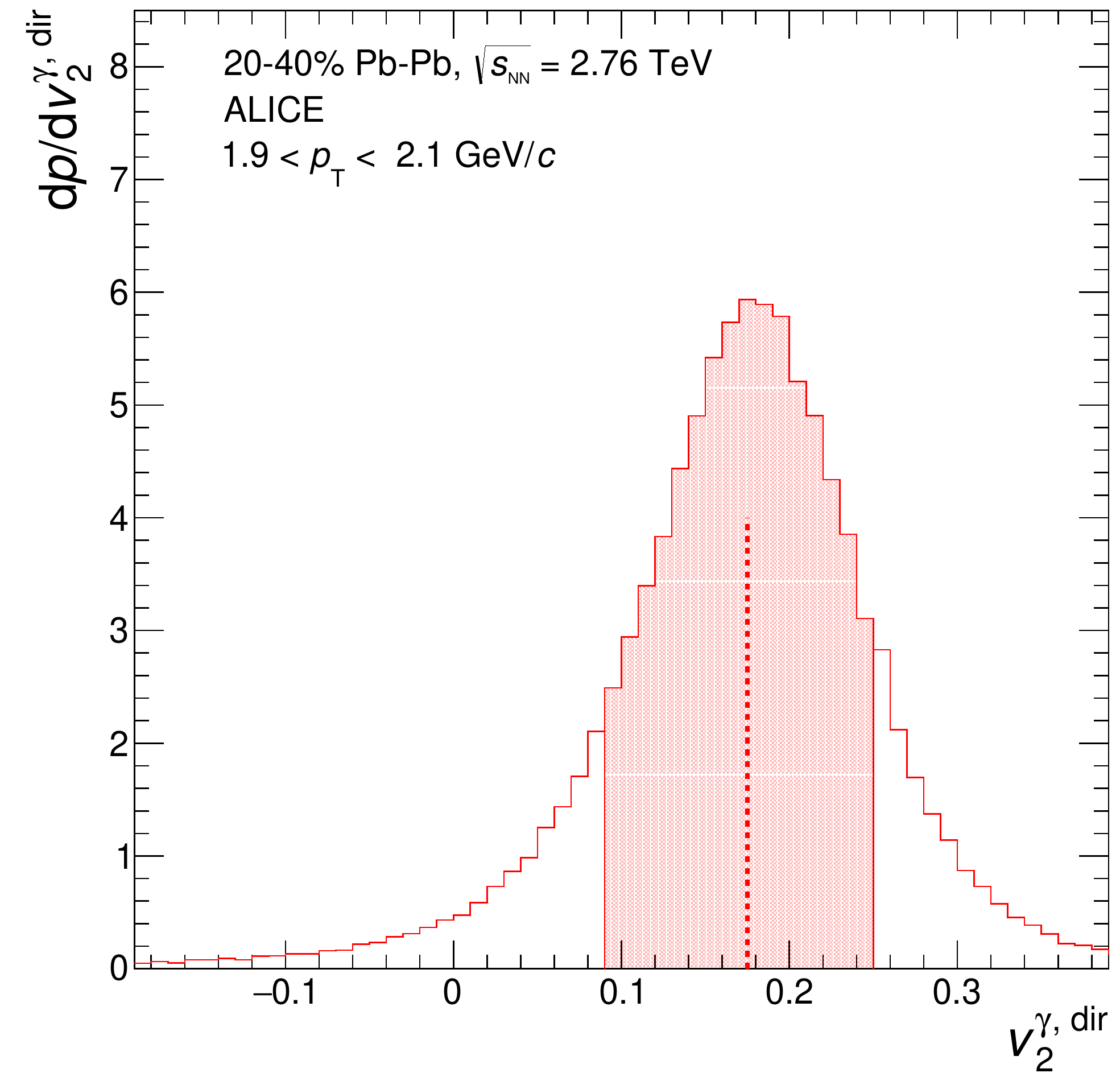}
\caption{ (Color online) Left: Central value (solid red line) and uncertainty of the direct-photon $v_2$ for a selected $\pT$ interval. The upper and lower edges of the red shaded area correspond to the total uncertainty of $\vdirg$ as obtained from linear Gaussian propagation of the uncertainties $\sigma(\vincg)$ and $\sigma(\vdecg)$. The Gaussian (with arbitrary normalization) reflects the measured value of $\Rg$ in this $\pT$ interval (blue dashed line) and its $\pm 1\sigma$ uncertainty (dark-blue shaded interval). Right: Posterior distribution of the true value of $\vdirg$ for the same interval in the Bayesian approach.
Note that the distribution has a non-Gaussian shape, implying that the $\pm$2$\sigma$ interval typically corresponds to a probability of less than 95.45\% as would be the case for a Gaussian.
\label{fig:RgammaTRUE}}
\end{figure}

The direct-photon $v_2$ is calculated from the combined PCM and PHOS photon excess $\Rg$ \cite{Adam:2015lda}, the combined inclusive $v_2$, and the calculated decay photon $v_2$. In the propagation of uncertainties, the relatively small significance of the photon excess of about 1--3 standard deviations (depending on the centrality class and $\pT$ interval) requires special attention. This is illustrated for a selected $\pT$ interval in the left panel of Fig.~\ref{fig:RgammaTRUE} which shows the obtained $\vdirg$ and its uncertainty as a function of the photon excess $\Rg$. The Gaussian function in this panel represents the measured value of $\Rg$ in this $\pT$ interval (dashed line) and its $1\sigma$ total uncertainty (dark blue shaded area). 
For $\Rg \lesssim 1.05$ one loses the sensitivity to $\vdirg$ as the uncertainties, indicated by the red shaded band, increase drastically. With the current uncertainties on $\Rg$ we cannot rule out completely that $\Rg \lesssim 1.05$.

We address the limited significance of the direct-photon excess by employing a Bayesian approach.
The parameters $R_\mathrm{\gamma,t}$, $v_2^\mathrm{\gamma,inc,t}$, $v_2^\mathrm{\gamma,dec,t}$ denoting the true values carry the index ``t'' and the measured quantities $R_\mathrm{\gamma,m}$, $v_2^\mathrm{\gamma,dec,m}$, $v_2^\mathrm{\gamma,dec,m}$ the index ``m''. Note that $R_\mathrm{\gamma,t}$ is restricted to its physically allowed range ($R_\mathrm{\gamma,t} \ge 1$), while the measured value $R_\mathrm{\gamma,m}$ can fluctuate below unity. The posterior distribution of the true parameters can be written as
\begin{equation}
P(\vec \vartheta | \vec m) \propto P( \vec m | \vec \vartheta) \pi(\vec \vartheta), \quad
\pi(\vec \vartheta) \equiv \pi(\vec R_\mathrm{\gamma,t}) = \Theta(R_{\mathrm{\gamma,t},1}-1, ..., R_{\mathrm{\gamma,t},n}-1) ,
\end{equation}
where in $\vec m = (\vec R_\mathrm{\gamma,m}, \vec v_2^\mathrm{\gamma,inc,m}, \vec v_2^\mathrm{\gamma,dec,m})$, $\vec \vartheta = (\vec R_\mathrm{\gamma,t}, \vec v_2^\mathrm{\gamma,inc,t}, \vec v_2^\mathrm{\gamma,dec,t})$. Here we use the notation introduced in Eq.~(\ref{v2incl}): vectors represent sets of measurements in different $\pT$ bins and $n$ is the number of these bins. The function $\pi(\vec R_\mathrm{\gamma,t})$ encodes the prior knowledge about $R_\gamma$. 
The multivariate Heaviside $\Theta$ function corresponds to a constant (improper) prior for $R_\mathrm{\gamma,t} \ge 1$. The probability to observe a certain set of measured values given the true values is modeled with multivariate Gaussians $G(\vec x;\,\vec \mu, V)$ (where $\vec \mu$ is the vector of mean values and $V$ is the covariance matrix):
\begin{equation}
P( \vec m | \vec \vartheta) = \prod_{x = R_\gamma,\,v_2^\mathrm{\gamma,inc},\,v_2^\mathrm{\gamma,dec}} G(\vec x_\mathrm{m};\, \vec x_\mathrm{t}, V_x).
\end{equation}
By sampling the posterior distribution $P(\vec \vartheta | \vec m)$, we obtain triplets $(\Rg, \vincg, \vdecg)$ for each $\pT$ bin from which we calculate $\vdirg$ according to Eq.~\ref{vdir}. An example of the resulting distribution for $\vdirg$ is shown in Fig.~\ref{fig:RgammaTRUE} (right panel). The medians of the $\vdirg$ distributions are taken as central values. The lower and upper edges of the error bars correspond to values of $v_2^\mathrm{dir}$ at which the integral of the $v_2$ distribution is 15.87\% and 84.13\% of the total integral. In case of a Gaussian distribution this corresponds to $1\sigma$ uncertainties.





\begin{figure}[ht]
\unitlength\textwidth
\centering
\includegraphics[width=0.49\linewidth]{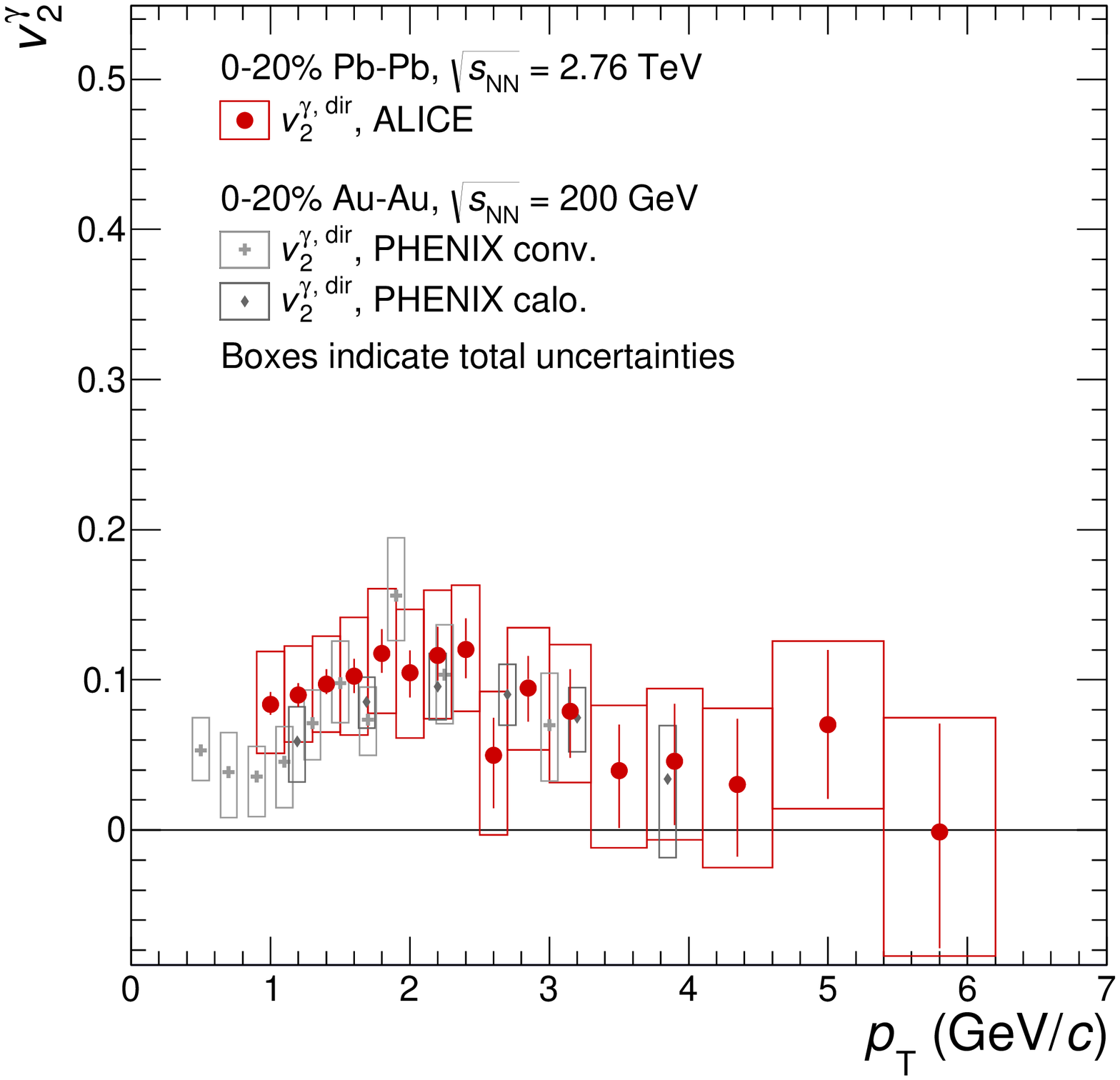}
\hfill
\includegraphics[width=0.49\linewidth]{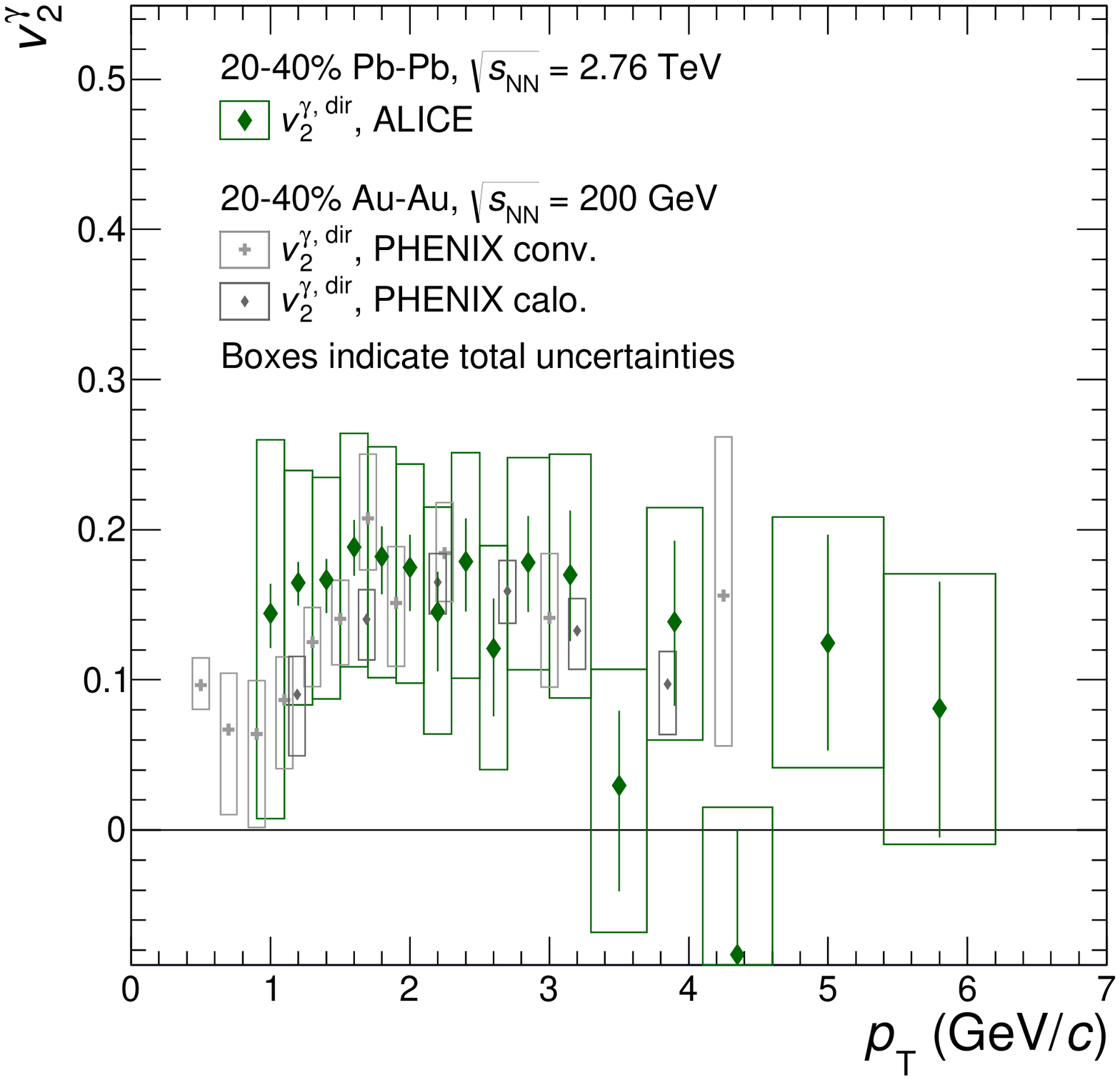}
\caption{(Color online) Elliptic flow of direct photons compared with PHENIX results \cite{Adare:2015lcd} for the 0--20\% (left) and 20--40\% (right) centrality classes. The vertical bars on each data point indicate the statistical uncertainties and the boxes the total uncertainty.
\label{fig:v2DirPHENIX}}
\end{figure}

The results for the direct-photon elliptic flow for the two centrality classes, 0--20\% and 20--40\%, are shown in Fig.~\ref{fig:v2DirPHENIX}. The total uncertainties, reflecting the Bayesian posterior distributions, are shown as boxes, and the error bars represent statistical uncertainties. The correlation of $v_\mathrm{2,dir}$ points for different $\pT$ bins as quantified by the correlation matrix is strong at low $\pT \lesssim \unit[2]{GeV}/c$ (correlation coefficients typically in the range 0.6--0.75) whereas the uncertainties at high $\pT$ are dominated by statistical uncertainties. We compare our results to measurements made at RHIC energies by the PHENIX collaboration \cite{Adare:2015lcd}. The inclusive photon $v_2$ was measured by PHENIX through the reconstruction of $e^+e^-$ pairs from photon conversions and with an electromagnetic calorimeter. The direct-photon elliptic flow in Au--Au collisions at RHIC and in Pb--Pb collisions at the LHC are found to be compatible within uncertainties. A simple explanation of the large and similar direct-photon elliptic flow for $\pT \lesssim \unit[2]{GeV}/c$ at RHIC and the LHC is that the bulk of the thermal direct photons is produced late at temperatures close to the transition temperature $T_\mathrm{c}$. This is interesting as na\"ively one would expect the $T^2$ temperature dependence of the photon emission rate to make the early hot QGP phase after thermalization also the brightest phase.

\begin{figure}[ht]
\unitlength\textwidth
\centering
\includegraphics[width=0.49\linewidth]{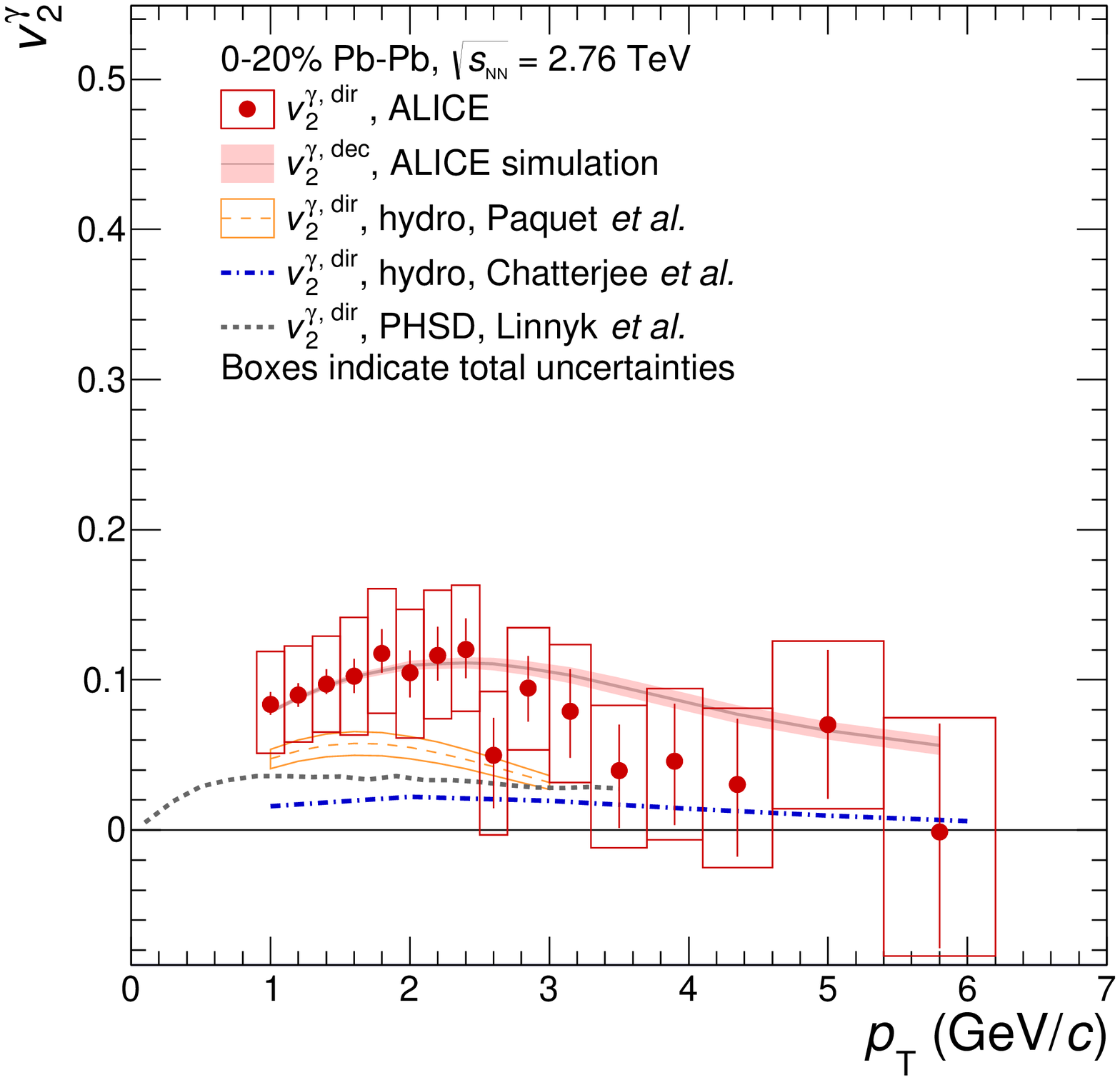}
\hfill
\includegraphics[width=0.49\linewidth]{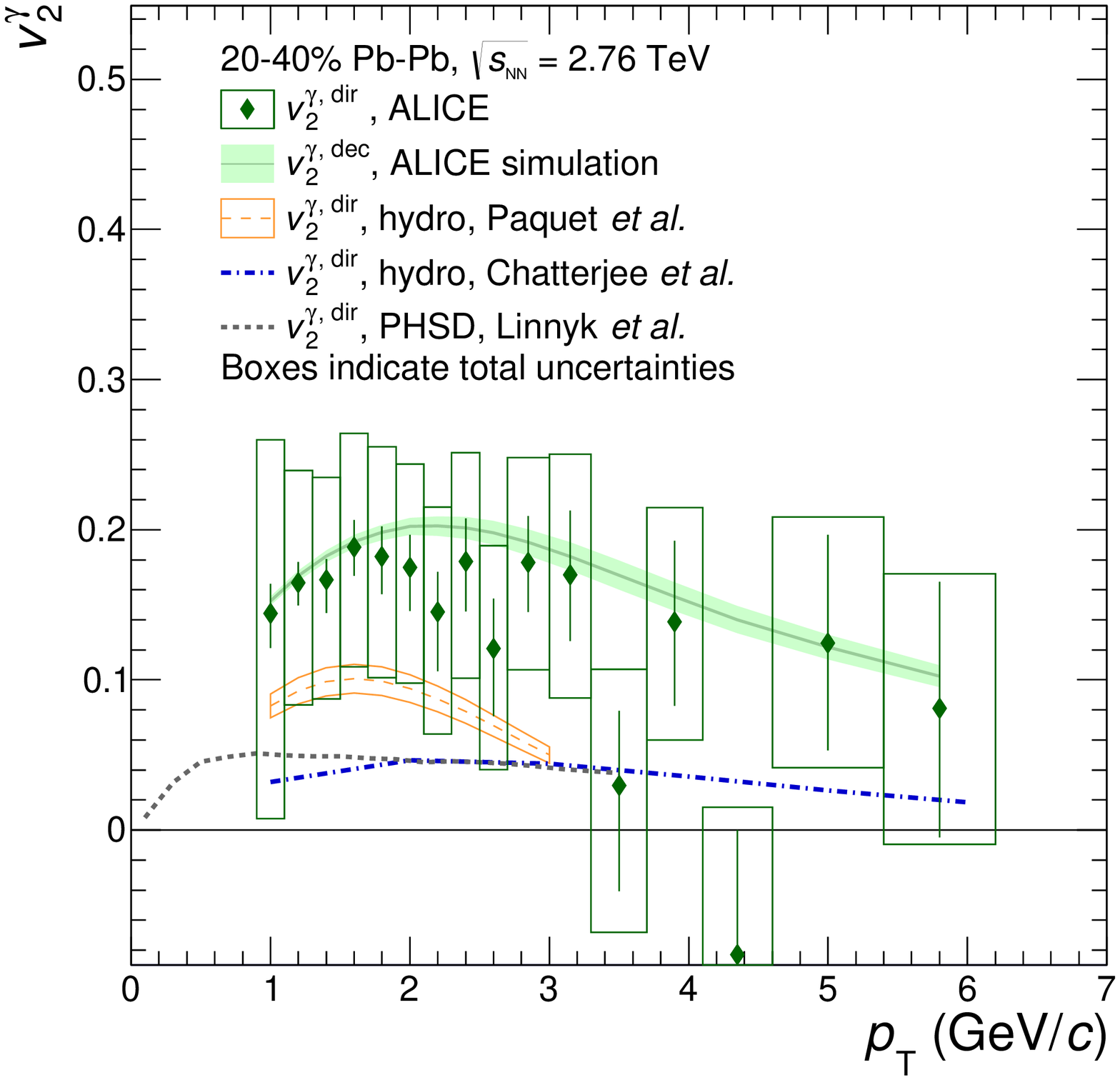}
\caption{(Color online) Elliptic flow of direct photons compared to model calculations in the 0--20\% (left) and 20--40\% (right) centrality classes. The vertical bars on each data point indicate the statistical uncertainties and the boxes the total uncertainty.
\label{fig:v2DirTHEORY} }
\end{figure}

Figure~\ref{fig:v2DirTHEORY} compares the measured direct-photon elliptic flow $\vdirg$ to the estimated decay photon elliptic flow $\vdecg$, marked as cocktail, and to the predictions of several theoretical models. Similarly to measurements at RHIC energies \cite{Adare:2011zr}, we find that the direct and decay photon elliptic flow are similar. 
We compare our measurements to state-of-the-art hydrodynamic model calculations \cite{Gale:2014dfa,Chatterjee:2017akg} and the PHSD transport model \cite{Linnyk:2015tha}.
The measured direct-photon elliptic flow is systematically higher than theoretical predictions, but is still compatible.

In order to quantify the deviation of the direct-photon $v_2$ measurement from a certain hypothesis with a frequentist $p$-value or, equivalently, the corresponding number of standard deviations, we use a Bayesian-inspired method \cite{Cousins:1991qz}. In this approach, the likelihood $L(\vec v_2^\mathrm{\gamma,inc,m}|\vec v_2^\mathrm{\gamma,dir,t})$ serves as a test statistic and is obtained by integrating over the nuisance parameters $\vec v_2^\mathrm{\gamma,dec,t}$ and $\vec R_{\gamma,t}$ using their Bayesian posterior distributions as weights. We focus on the interval $0.9 < \pT < 2.1~\mathrm{GeV}/c$ in which the contribution of thermal photons is expected to be important. The significance of the deviation from the hypothesis $v_2^\mathrm{\gamma,dir,t} = 0$ for individual $\pT$ bins is in the range 1.8--2.1$\sigma$ for the 0--20\% class and 0.9--1.5$\sigma$ for the 20--40\% class. We also go a step further and estimate the combined significance of the deviation from the hypothesis $\vdirg \equiv 0$ for this $\pT$ interval. This tests in addition how well the shape of $v_2^\mathrm{\gamma,inc,m}$ as a function of $\pT$ agrees with $v_2^\mathrm{\gamma,dec,m} / R_\gamma$, i.e., with the expectation for $\vdirg \equiv 0$. We estimate the covariance matrix describing the correlation by characterizing the different sources of systematic uncertainties of $R_\gamma$, the inclusive, and the decay photon flow as either fully uncorrelated or fully correlated in $\pT$. Varying the assumptions about the correlation of the data points we obtain significances of typically less than $1\sigma$ for both centrality classes. While the applied method is essential for a meaningful comparison of the $v_2^\mathrm{\gamma,dir}$ data with different model predictions, the methods to estimate the covariance matrix can be improved in future analyses. 


\section{Conclusions}

In summary, we report the first measurement of elliptic flow of inclusive and direct photons as a function of transverse momentum in the range $0.9<\pT<6.2$~GeV/$c$ for central and semi-central Pb--Pb collisions at $\snn=2.76$ TeV. 
The elliptic flow of inclusive photons was measured with the scalar product method, independently in the electromagnetic calorimeter PHOS and with the photon conversion method where the reference particles in both cases were measured by the V0--A and V0--C detectors.
The combined inclusive photon $\vincg$, together with the calculated decay photon $\vdecg$ and the previously measured $\Rg$ are used to calculate the elliptic flow of direct photons. The measured direct-photon flow $\vdirg$ appears to be close to the decay photon flow for both centrality classes, similar to observations at lower collision energies. Moreover, the measured $\vdirg$ is similar to the measurements by the PHENIX collaboration at RHIC. The considered hydrodynamic and transport models predict a larger inclusive photon elliptic flow (by approximately $40$\%) and a smaller direct-photon elliptic flow than observed. With current uncertainties, however, these models are consistent with the presented direct-photon elliptic flow data. Future measurements using a larger statistics dataset will greatly increase the precision of this measurement and allow us to extend the measurement to higher $\pT$, since the statistical uncertainty is dominating the total uncertainty for $\pT>2.0$~GeV/$c$ and $\pT>3.0$~GeV/$c$ for the PHOS and PCM inclusive photon flow measurement, respectively. In addition, a larger statistics dataset will also help to constrain the systematic uncertainties on the inclusive and decay photon flow, as well as the measurement of $\Rg$ over the whole $\pT$ range. A further reduction of the systematic uncertainties is expected from improved detector knowledge. For instance, in case of PCM the largest systematic uncertainty in the measurement of $\Rg$ is related to modeling the material in which the photons convert. Calibrating regions of the detector with less well known material budget based on regions with very well known material might significantly reduce the overall material budget uncertainty. The $\Rg$ measurement can be improved further by measuring neutral pion and eta meson spectra in a combined PCM-calorimeter approach in which one decay photon is measured through conversion and the other with a calorimeter.

\newenvironment{acknowledgement}{\relax}{\relax}
\begin{acknowledgement}
\section*{Acknowledgements}
We would like to thank 
Elena Bratkovskaya and
Jean-Fran\c{c}ois Paquet
for providing calculations shown in this paper and for useful discussions.


The ALICE Collaboration would like to thank all its engineers and technicians for their invaluable contributions to the construction of the experiment and the CERN accelerator teams for the outstanding performance of the LHC complex.
The ALICE Collaboration gratefully acknowledges the resources and support provided by all Grid centres and the Worldwide LHC Computing Grid (WLCG) collaboration.
The ALICE Collaboration acknowledges the following funding agencies for their support in building and running the ALICE detector:
A. I. Alikhanyan National Science Laboratory (Yerevan Physics Institute) Foundation (ANSL), State Committee of Science and World Federation of Scientists (WFS), Armenia;
Austrian Academy of Sciences and Nationalstiftung f\"{u}r Forschung, Technologie und Entwicklung, Austria;
Ministry of Communications and High Technologies, National Nuclear Research Center, Azerbaijan;
Conselho Nacional de Desenvolvimento Cient\'{\i}fico e Tecnol\'{o}gico (CNPq), Universidade Federal do Rio Grande do Sul (UFRGS), Financiadora de Estudos e Projetos (Finep) and Funda\c{c}\~{a}o de Amparo \`{a} Pesquisa do Estado de S\~{a}o Paulo (FAPESP), Brazil;
Ministry of Science \& Technology of China (MSTC), National Natural Science Foundation of China (NSFC) and Ministry of Education of China (MOEC) , China;
Ministry of Science and Education, Croatia;
Ministry of Education, Youth and Sports of the Czech Republic, Czech Republic;
The Danish Council for Independent Research | Natural Sciences, the Carlsberg Foundation and Danish National Research Foundation (DNRF), Denmark;
Helsinki Institute of Physics (HIP), Finland;
Commissariat \`{a} l'Energie Atomique (CEA) and Institut National de Physique Nucl\'{e}aire et de Physique des Particules (IN2P3) and Centre National de la Recherche Scientifique (CNRS), France;
Bundesministerium f\"{u}r Bildung, Wissenschaft, Forschung und Technologie (BMBF) and GSI Helmholtzzentrum f\"{u}r Schwerionenforschung GmbH, Germany;
General Secretariat for Research and Technology, Ministry of Education, Research and Religions, Greece;
National Research, Development and Innovation Office, Hungary;
Department of Atomic Energy Government of India (DAE), Department of Science and Technology, Government of India (DST), University Grants Commission, Government of India (UGC) and Council of Scientific and Industrial Research (CSIR), India;
Indonesian Institute of Science, Indonesia;
Centro Fermi - Museo Storico della Fisica e Centro Studi e Ricerche Enrico Fermi and Istituto Nazionale di Fisica Nucleare (INFN), Italy;
Institute for Innovative Science and Technology , Nagasaki Institute of Applied Science (IIST), Japan Society for the Promotion of Science (JSPS) KAKENHI and Japanese Ministry of Education, Culture, Sports, Science and Technology (MEXT), Japan;
Consejo Nacional de Ciencia (CONACYT) y Tecnolog\'{i}a, through Fondo de Cooperaci\'{o}n Internacional en Ciencia y Tecnolog\'{i}a (FONCICYT) and Direcci\'{o}n General de Asuntos del Personal Academico (DGAPA), Mexico;
Nederlandse Organisatie voor Wetenschappelijk Onderzoek (NWO), Netherlands;
The Research Council of Norway, Norway;
Commission on Science and Technology for Sustainable Development in the South (COMSATS), Pakistan;
Pontificia Universidad Cat\'{o}lica del Per\'{u}, Peru;
Ministry of Science and Higher Education and National Science Centre, Poland;
Korea Institute of Science and Technology Information and National Research Foundation of Korea (NRF), Republic of Korea;
Ministry of Education and Scientific Research, Institute of Atomic Physics and Romanian National Agency for Science, Technology and Innovation, Romania;
Joint Institute for Nuclear Research (JINR), Ministry of Education and Science of the Russian Federation and National Research Centre Kurchatov Institute, Russia;
Ministry of Education, Science, Research and Sport of the Slovak Republic, Slovakia;
National Research Foundation of South Africa, South Africa;
Centro de Aplicaciones Tecnol\'{o}gicas y Desarrollo Nuclear (CEADEN), Cubaenerg\'{\i}a, Cuba and Centro de Investigaciones Energ\'{e}ticas, Medioambientales y Tecnol\'{o}gicas (CIEMAT), Spain;
Swedish Research Council (VR) and Knut \& Alice Wallenberg Foundation (KAW), Sweden;
European Organization for Nuclear Research, Switzerland;
National Science and Technology Development Agency (NSDTA), Suranaree University of Technology (SUT) and Office of the Higher Education Commission under NRU project of Thailand, Thailand;
Turkish Atomic Energy Agency (TAEK), Turkey;
National Academy of  Sciences of Ukraine, Ukraine;
Science and Technology Facilities Council (STFC), United Kingdom;
National Science Foundation of the United States of America (NSF) and United States Department of Energy, Office of Nuclear Physics (DOE NP), United States of America.    
\end{acknowledgement}

\ifcernpp
\bibliographystyle{utphys} 
\else
\bibliographystyle{h-physrev}
\fi
\bibliography{biblio}

\newpage
\appendix
\section{The ALICE Collaboration}
\label{app:collab}

\begingroup
\small
\begin{flushleft}
S.~Acharya\Irefn{org139}\And 
F.T.-.~Acosta\Irefn{org21}\And 
D.~Adamov\'{a}\Irefn{org94}\And 
J.~Adolfsson\Irefn{org81}\And 
M.M.~Aggarwal\Irefn{org98}\And 
G.~Aglieri Rinella\Irefn{org35}\And 
M.~Agnello\Irefn{org32}\And 
N.~Agrawal\Irefn{org49}\And 
Z.~Ahammed\Irefn{org139}\And 
S.U.~Ahn\Irefn{org77}\And 
S.~Aiola\Irefn{org144}
% ...
\And
\renewcommand\labelenumi{\textsuperscript{\theenumi}~}

\section*{Affiliation notes}
\renewcommand\theenumi{\roman{enumi}}
\begin{Authlist}
\item \Adef{org*}Deceased
\item \Adef{orgI}Dipartimento DET del Politecnico di Torino, Turin, Italy
\item \Adef{orgII}M.V. Lomonosov Moscow State University, D.V. Skobeltsyn Institute of Nuclear, Physics, Moscow, Russia
\item \Adef{orgIII}Department of Applied Physics, Aligarh Muslim University, Aligarh, India
\item \Adef{orgIV}Institute of Theoretical Physics, University of Wroclaw, Poland
\end{Authlist}

\section*{Collaboration Institutes}
\renewcommand\theenumi{\arabic{enumi}~}
\begin{Authlist}
\item \Idef{org1}A.I. Alikhanyan National Science Laboratory (Yerevan Physics Institute) Foundation, Yerevan, Armenia
\item \Idef{org2}Bogolyubov Institute for Theoretical Physics, National Academy of Sciences of Ukraine, Kiev, Ukraine
\item \Idef{org3}Bose Institute, Department of Physics  and Centre for Astroparticle Physics and Space Science (CAPSS), Kolkata, India
\item \Idef{org4}Budker Institute for Nuclear Physics, Novosibirsk, Russia
\item \Idef{org5}California Polytechnic State University, San Luis Obispo, California, United States
\item \Idef{org6}Central China Normal University, Wuhan, China
\item \Idef{org7}Centre de Calcul de l'IN2P3, Villeurbanne, Lyon, France
\item \Idef{org8}Centro de Aplicaciones Tecnol\'{o}gicas y Desarrollo Nuclear (CEADEN), Havana, Cuba
\item \Idef{org9}Centro de Investigaci\'{o}n y de Estudios Avanzados (CINVESTAV), Mexico City and M\'{e}rida, Mexico
\item \Idef{org10}Centro Fermi - Museo Storico della Fisica e Centro Studi e Ricerche ``Enrico Fermi', Rome, Italy
\item \Idef{org11}Chicago State University, Chicago, Illinois, United States
\item \Idef{org12}China Institute of Atomic Energy, Beijing, China
\item \Idef{org13}Chonbuk National University, Jeonju, Republic of Korea
\item \Idef{org14}Comenius University Bratislava, Faculty of Mathematics, Physics and Informatics, Bratislava, Slovakia
\item \Idef{org15}COMSATS Institute of Information Technology (CIIT), Islamabad, Pakistan
\item \Idef{org16}Creighton University, Omaha, Nebraska, United States
\item \Idef{org17}Department of Physics, Aligarh Muslim University, Aligarh, India
\item \Idef{org18}Department of Physics, Ohio State University, Columbus, Ohio, United States
\item \Idef{org19}Department of Physics, Pusan National University, Pusan, Republic of Korea
\item \Idef{org20}Department of Physics, Sejong University, Seoul, Republic of Korea
\item \Idef{org21}Department of Physics, University of California, Berkeley, California, United States
\item \Idef{org22}Department of Physics, University of Oslo, Oslo, Norway
\item \Idef{org23}Department of Physics and Technology, University of Bergen, Bergen, Norway
\item \Idef{org24}Dipartimento di Fisica dell'Universit\`{a} 'La Sapienza' and Sezione INFN, Rome, Italy
\item \Idef{org25}Dipartimento di Fisica dell'Universit\`{a} and Sezione INFN, Cagliari, Italy
\item \Idef{org26}Dipartimento di Fisica dell'Universit\`{a} and Sezione INFN, Trieste, Italy
\item \Idef{org27}Dipartimento di Fisica dell'Universit\`{a} and Sezione INFN, Turin, Italy
\item \Idef{org28}Dipartimento di Fisica e Astronomia dell'Universit\`{a} and Sezione INFN, Bologna, Italy
\item \Idef{org29}Dipartimento di Fisica e Astronomia dell'Universit\`{a} and Sezione INFN, Catania, Italy
\item \Idef{org30}Dipartimento di Fisica e Astronomia dell'Universit\`{a} and Sezione INFN, Padova, Italy
\item \Idef{org31}Dipartimento di Fisica `E.R.~Caianiello' dell'Universit\`{a} and Gruppo Collegato INFN, Salerno, Italy
\item \Idef{org32}Dipartimento DISAT del Politecnico and Sezione INFN, Turin, Italy
\item \Idef{org33}Dipartimento di Scienze e Innovazione Tecnologica dell'Universit\`{a} del Piemonte Orientale and INFN Sezione di Torino, Alessandria, Italy
\item \Idef{org34}Dipartimento Interateneo di Fisica `M.~Merlin' and Sezione INFN, Bari, Italy
\item \Idef{org35}European Organization for Nuclear Research (CERN), Geneva, Switzerland
\item \Idef{org36}Faculty of Electrical Engineering, Mechanical Engineering and Naval Architecture, University of Split, Split, Croatia
\item \Idef{org37}Faculty of Engineering and Science, Western Norway University of Applied Sciences, Bergen, Norway
\item \Idef{org38}Faculty of Nuclear Sciences and Physical Engineering, Czech Technical University in Prague, Prague, Czech Republic
\item \Idef{org39}Faculty of Science, P.J.~\v{S}af\'{a}rik University, Ko\v{s}ice, Slovakia
\item \Idef{org40}Frankfurt Institute for Advanced Studies, Johann Wolfgang Goethe-Universit\"{a}t Frankfurt, Frankfurt, Germany
\item \Idef{org41}Gangneung-Wonju National University, Gangneung, Republic of Korea
\item \Idef{org42}Gauhati University, Department of Physics, Guwahati, India
\item \Idef{org43}Helmholtz-Institut f\"{u}r Strahlen- und Kernphysik, Rheinische Friedrich-Wilhelms-Universit\"{a}t Bonn, Bonn, Germany
\item \Idef{org44}Helsinki Institute of Physics (HIP), Helsinki, Finland
\item \Idef{org45}High Energy Physics Group,  Universidad Aut\'{o}noma de Puebla, Puebla, Mexico
\item \Idef{org46}Hiroshima University, Hiroshima, Japan
\item \Idef{org47}Hochschule Worms, Zentrum  f\"{u}r Technologietransfer und Telekommunikation (ZTT), Worms, Germany
\item \Idef{org48}Horia Hulubei National Institute of Physics and Nuclear Engineering, Bucharest, Romania
\item \Idef{org49}Indian Institute of Technology Bombay (IIT), Mumbai, India
\item \Idef{org50}Indian Institute of Technology Indore, Indore, India
\item \Idef{org51}Indonesian Institute of Sciences, Jakarta, Indonesia
\item \Idef{org52}INFN, Laboratori Nazionali di Frascati, Frascati, Italy
\item \Idef{org53}INFN, Sezione di Bari, Bari, Italy
\item \Idef{org54}INFN, Sezione di Bologna, Bologna, Italy
\item \Idef{org55}INFN, Sezione di Cagliari, Cagliari, Italy
\item \Idef{org56}INFN, Sezione di Catania, Catania, Italy
\item \Idef{org57}INFN, Sezione di Padova, Padova, Italy
\item \Idef{org58}INFN, Sezione di Roma, Rome, Italy
\item \Idef{org59}INFN, Sezione di Torino, Turin, Italy
\item \Idef{org60}INFN, Sezione di Trieste, Trieste, Italy
\item \Idef{org61}Inha University, Incheon, Republic of Korea
\item \Idef{org62}Institut de Physique Nucl\'{e}aire d'Orsay (IPNO), Institut National de Physique Nucl\'{e}aire et de Physique des Particules (IN2P3/CNRS), Universit\'{e} de Paris-Sud, Universit\'{e} Paris-Saclay, Orsay, France
\item \Idef{org63}Institute for Nuclear Research, Academy of Sciences, Moscow, Russia
\item \Idef{org64}Institute for Subatomic Physics, Utrecht University/Nikhef, Utrecht, Netherlands
\item \Idef{org65}Institute for Theoretical and Experimental Physics, Moscow, Russia
\item \Idef{org66}Institute of Experimental Physics, Slovak Academy of Sciences, Ko\v{s}ice, Slovakia
\item \Idef{org67}Institute of Physics, Homi Bhabha National Institute, Bhubaneswar, India
\item \Idef{org68}Institute of Physics of the Czech Academy of Sciences, Prague, Czech Republic
\item \Idef{org69}Institute of Space Science (ISS), Bucharest, Romania
\item \Idef{org70}Institut f\"{u}r Kernphysik, Johann Wolfgang Goethe-Universit\"{a}t Frankfurt, Frankfurt, Germany
\item \Idef{org71}Instituto de Ciencias Nucleares, Universidad Nacional Aut\'{o}noma de M\'{e}xico, Mexico City, Mexico
\item \Idef{org72}Instituto de F\'{i}sica, Universidade Federal do Rio Grande do Sul (UFRGS), Porto Alegre, Brazil
\item \Idef{org73}Instituto de F\'{\i}sica, Universidad Nacional Aut\'{o}noma de M\'{e}xico, Mexico City, Mexico
\item \Idef{org74}iThemba LABS, National Research Foundation, Somerset West, South Africa
\item \Idef{org75}Johann-Wolfgang-Goethe Universit\"{a}t Frankfurt Institut f\"{u}r Informatik, Fachbereich Informatik und Mathematik, Frankfurt, Germany
\item \Idef{org76}Joint Institute for Nuclear Research (JINR), Dubna, Russia
\item \Idef{org77}Korea Institute of Science and Technology Information, Daejeon, Republic of Korea
\item \Idef{org78}KTO Karatay University, Konya, Turkey
\item \Idef{org79}Laboratoire de Physique Subatomique et de Cosmologie, Universit\'{e} Grenoble-Alpes, CNRS-IN2P3, Grenoble, France
\item \Idef{org80}Lawrence Berkeley National Laboratory, Berkeley, California, United States
\item \Idef{org81}Lund University Department of Physics, Division of Particle Physics, Lund, Sweden
\item \Idef{org82}Nagasaki Institute of Applied Science, Nagasaki, Japan
\item \Idef{org83}Nara Women{'}s University (NWU), Nara, Japan
\item \Idef{org84}National and Kapodistrian University of Athens, School of Science, Department of Physics , Athens, Greece
\item \Idef{org85}National Centre for Nuclear Research, Warsaw, Poland
\item \Idef{org86}National Institute of Science Education and Research, Homi Bhabha National Institute, Jatni, India
\item \Idef{org87}National Nuclear Research Center, Baku, Azerbaijan
\item \Idef{org88}National Research Centre Kurchatov Institute, Moscow, Russia
\item \Idef{org89}Niels Bohr Institute, University of Copenhagen, Copenhagen, Denmark
\item \Idef{org90}Nikhef, National institute for subatomic physics, Amsterdam, Netherlands
\item \Idef{org91}NRC Kurchatov Institute IHEP, Protvino, Russia
\item \Idef{org92}NRNU Moscow Engineering Physics Institute, Moscow, Russia
\item \Idef{org93}Nuclear Physics Group, STFC Daresbury Laboratory, Daresbury, United Kingdom
\item \Idef{org94}Nuclear Physics Institute of the Czech Academy of Sciences, \v{R}e\v{z} u Prahy, Czech Republic
\item \Idef{org95}Oak Ridge National Laboratory, Oak Ridge, Tennessee, United States
\item \Idef{org96}Petersburg Nuclear Physics Institute, Gatchina, Russia
\item \Idef{org97}Physics department, Faculty of science, University of Zagreb, Zagreb, Croatia
\item \Idef{org98}Physics Department, Panjab University, Chandigarh, India
\item \Idef{org99}Physics Department, University of Jammu, Jammu, India
\item \Idef{org100}Physics Department, University of Rajasthan, Jaipur, India
\item \Idef{org101}Physikalisches Institut, Eberhard-Karls-Universit\"{a}t T\"{u}bingen, T\"{u}bingen, Germany
\item \Idef{org102}Physikalisches Institut, Ruprecht-Karls-Universit\"{a}t Heidelberg, Heidelberg, Germany
\item \Idef{org103}Physik Department, Technische Universit\"{a}t M\"{u}nchen, Munich, Germany
\item \Idef{org104}Research Division and ExtreMe Matter Institute EMMI, GSI Helmholtzzentrum f\"ur Schwerionenforschung GmbH, Darmstadt, Germany
\item \Idef{org105}Rudjer Bo\v{s}kovi\'{c} Institute, Zagreb, Croatia
\item \Idef{org106}Russian Federal Nuclear Center (VNIIEF), Sarov, Russia
\item \Idef{org107}Saha Institute of Nuclear Physics, Homi Bhabha National Institute, Kolkata, India
\item \Idef{org108}School of Physics and Astronomy, University of Birmingham, Birmingham, United Kingdom
\item \Idef{org109}Secci\'{o}n F\'{\i}sica, Departamento de Ciencias, Pontificia Universidad Cat\'{o}lica del Per\'{u}, Lima, Peru
\item \Idef{org110}Shanghai Institute of Applied Physics, Shanghai, China
\item \Idef{org111}St. Petersburg State University, St. Petersburg, Russia
\item \Idef{org112}Stefan Meyer Institut f\"{u}r Subatomare Physik (SMI), Vienna, Austria
\item \Idef{org113}SUBATECH, IMT Atlantique, Universit\'{e} de Nantes, CNRS-IN2P3, Nantes, France
\item \Idef{org114}Suranaree University of Technology, Nakhon Ratchasima, Thailand
\item \Idef{org115}Technical University of Ko\v{s}ice, Ko\v{s}ice, Slovakia
\item \Idef{org116}Technische Universit\"{a}t M\"{u}nchen, Excellence Cluster 'Universe', Munich, Germany
\item \Idef{org117}The Henryk Niewodniczanski Institute of Nuclear Physics, Polish Academy of Sciences, Cracow, Poland
\item \Idef{org118}The University of Texas at Austin, Austin, Texas, United States
\item \Idef{org119}Universidad Aut\'{o}noma de Sinaloa, Culiac\'{a}n, Mexico
\item \Idef{org120}Universidade de S\~{a}o Paulo (USP), S\~{a}o Paulo, Brazil
\item \Idef{org121}Universidade Estadual de Campinas (UNICAMP), Campinas, Brazil
\item \Idef{org122}Universidade Federal do ABC, Santo Andre, Brazil
\item \Idef{org123}University College of Southeast Norway, Tonsberg, Norway
\item \Idef{org124}University of Cape Town, Cape Town, South Africa
\item \Idef{org125}University of Houston, Houston, Texas, United States
\item \Idef{org126}University of Jyv\"{a}skyl\"{a}, Jyv\"{a}skyl\"{a}, Finland
\item \Idef{org127}University of Liverpool, Liverpool, United Kingdom
\item \Idef{org128}University of Tennessee, Knoxville, Tennessee, United States
\item \Idef{org129}University of the Witwatersrand, Johannesburg, South Africa
\item \Idef{org130}University of Tokyo, Tokyo, Japan
\item \Idef{org131}University of Tsukuba, Tsukuba, Japan
\item \Idef{org132}Universit\'{e} Clermont Auvergne, CNRS/IN2P3, LPC, Clermont-Ferrand, France
\item \Idef{org133}Universit\'{e} de Lyon, Universit\'{e} Lyon 1, CNRS/IN2P3, IPN-Lyon, Villeurbanne, Lyon, France
\item \Idef{org134}Universit\'{e} de Strasbourg, CNRS, IPHC UMR 7178, F-67000 Strasbourg, France, Strasbourg, France
\item \Idef{org135} Universit\'{e} Paris-Saclay Centre d¿\'Etudes de Saclay (CEA), IRFU, Department de Physique Nucl\'{e}aire (DPhN), Saclay, France
\item \Idef{org136}Universit\`{a} degli Studi di Foggia, Foggia, Italy
\item \Idef{org137}Universit\`{a} degli Studi di Pavia, Pavia, Italy
\item \Idef{org138}Universit\`{a} di Brescia, Brescia, Italy
\item \Idef{org139}Variable Energy Cyclotron Centre, Homi Bhabha National Institute, Kolkata, India
\item \Idef{org140}Warsaw University of Technology, Warsaw, Poland
\item \Idef{org141}Wayne State University, Detroit, Michigan, United States
\item \Idef{org142}Westf\"{a}lische Wilhelms-Universit\"{a}t M\"{u}nster, Institut f\"{u}r Kernphysik, M\"{u}nster, Germany
\item \Idef{org143}Wigner Research Centre for Physics, Hungarian Academy of Sciences, Budapest, Hungary
\item \Idef{org144}Yale University, New Haven, Connecticut, United States
\item \Idef{org145}Yonsei University, Seoul, Republic of Korea
\end{Authlist}
\endgroup
\end{document}